\documentclass[11pt]{amsart}
\usepackage[cp1251]{inputenc}
\usepackage{graphicx,color}
\usepackage[left=1in,right=1in,top=1in,bottom=1in]{geometry}
\usepackage{latexsym}
\usepackage{cite}
\usepackage{amsmath,amssymb,amsthm,mathrsfs,amscd}

\newtheorem*{defi}{Definition}

\allowdisplaybreaks[1]  

\newtheorem{theorem}{Theorem}
\newtheorem{prop}[theorem]{Proposition}

\newtheorem{lemma}[theorem]{Lemma}
\newtheorem{defn}[theorem]{Definition}
\newtheorem{rem}[theorem]{Remark}

\renewcommand{\phi}{\varphi}

\let\epsilon=\varepsilon

\newcommand{\A}{\mathcal{A}}

\def\crn#1#2{{\vcenter{\vbox{
\hbox{\kern#2pt \vrule width.#2pt height#1pt
 }
\hrule height.#2pt}}}}

\newcounter{mnotecount}[section]
\let\oldmarginpar\marginpar
\renewcommand\marginpar[1]{\-\oldmarginpar[\raggedleft\footnotesize #1]%
{\raggedright\footnotesize #1}}

\begin{document}

%
%
%
%
%
%
%
%
%

\title{A comparison theorem for cosmological lightcones}

\author[M. Carfora]{Mauro Carfora ${}^{(\star)}$}

\address[Department of Physics, University of Pavia]{University of Pavia} 
\address[GNFM and INFN]{Italian National Group of Mathematical Physics, and INFN Pavia Section}
\email{mauro.carfora@unipv.it        (Corresponding author)}

\author[F. Familiari]{Francesca Familiari}

\address[Department of Physics, University of Pavia]{University of Pavia} 
\address[GNFM and INFN]{Italian National Group of Mathematical Physics, and INFN Pavia Section} 

\email{francesca.familiari01@universitadipavia.it}



\date{7 December 2020}

\begin{abstract}
Let $(M, g)$ denote a cosmological spacetime  describing the evolution of a universe which is isotropic and homogeneous on large scales, but highly inhomogeneous on smaller scales. We consider  two past lightcones, the first, $\mathcal{C}^-_L(p, g)$, is associated with the physical observer $p\in\,M$ who describes the actual physical spacetime geometry of   $(M, g)$ at the length scale $L$, whereas  the second, $\mathcal{C}^-_L(p, \hat{g})$, is associated with an idealized version of the observer $p$ who, notwithstanding the presence of local inhomogeneities at the given scale $L$, wish to model  $(M, g)$ with a member $(M, \hat{g})$  of the family of  Friedmann-Lemaitre-Robertson-Walker spacetimes. In such a framework, we discuss a number of mathematical results that allows a rigorous comparison between  the two lightcones   $\mathcal{C}^-_L(p, g)$ and $\mathcal{C}^-_L(p, \hat{g})$. In particular, we introduce a scale dependent ($L$) lightcone-comparison    functional, defined by a  harmonic type  energy, associated with a natural map between the physical $\mathcal{C}^-_L(p, g)$  and the FLRW reference lightcone  $\mathcal{C}^-_L(p, \hat{g})$. This functional has a number of remarkable properties, in particular  it vanishes 
iff, at the given length-scale, the corresponding lightcone surface sections (the celestial spheres)  are isometric. We discuss in detail its variational analysis and prove the existence of a minimum that characterizes a natural scale-dependent distance functional between the two lightcones.  We also indicate how it is possible to extend our results to the case when  caustics develop on the physical past lightcone $\mathcal{C}^-_L(p, g)$.   Finally, we show how the distance functional is related to spacetime scalar curvature in the causal past of the two lightcones, and briefly illustrate a number of  its possible applications.
\end{abstract}

\maketitle

\section{INTRODUCTION}
\label{Introduction}

\noindent

 For many among us, the first encounter with Boris Dubrovin has been through his classic books  \textit{G$\acute{e}$ometrie Contemporaine: m$\acute{e}$thodes et applications}, coauthored with S. Novikov and A. Fomenko,  and published by MIR in a silk-bonded three-volume set, an edition that, notwithstanding the later expanded Springer version \cite{Dubrovin}, we treasured with care. To the best of our knowledge Boris did not work in general relativity but in his \textit{G$\acute{e}$ometrie Contemporaine} there are  two wonderful little chapters, just short of a total of fifty  tersely written pages, which provide the fastest way to acquaint yourself with general relativity.  Thinking of these elegant pages,  we hope that it is not inappropriate to dedicate to Boris the present work addressing  a long-standing problem in mathematical  cosmology.\\
\\
 Let us recall that the observed universe is described by a spacetime $(M, g)$, a 4-dimensional manifold $M$ endowed with a Lorentzian metric $g$,  which is (statistically) isotropic and homogeneous only on sufficiently large scales, say $L\,\geq\,L_0$, where  the current acceptable figure for the homogeneity scale is $L_0\geq 100h^{-1}\,Mpc$, where $h$ is the dimensionless Hubble parameter  describing  the relative uncertainty of the true value of the present-epoch  Hubble-Lemaitre constant $H_0\,=\,100\,h\,\,Km/s/Mpc$. At these homogeneity scales $(M, g)$ is described with great accuracy by a member of the homogeneous and isotropic family of  Friedman--Lemaitre--Robertson--Walker (FLRW) spacetimes $(M, \hat{g})$.  At smaller scales, where inhomogeneities statistically dominate, we should resort to the full-fledged spacetime geometry of    $(M, g)$ in order to   provide the correct dynamical description of cosmological observations. However, coming to mathematical terms with the geometrical and physical structure of  $(M, g)$  is a daunting task  and typically we  keep on modeling the dynamics of the universe over these inhomogeneity scales with the  FLRW model $(M, \hat{g})$, thought of as providing a background  around which  the actual spacetime geometry $(M, g)$ is perturbatively expanded.  If we want to go beyond perturbation theory, we face the mathematically delicate problem of finding a way for  comparing the past lightcone region\footnote{Details on the notation adopted are explained in full detail in the body of the paper.  In particular the regions $\mathcal{C}^-_L(p, {g})$ and  $\mathcal{C}^-_L(p, \hat{g})$  are defined by (\ref{ConesRegions}).}    
 $\mathcal{C}^-_L(p, {g})$, associated with an (instantaneous) observer $p$,  sampling the inhomogeneities in $(M, g)$ at the given length scale $L$,  with the corresponding past lightcone region $\mathcal{C}^-_L(p, \hat{g})$ in the assumed FLRW background  $(M, \hat{g})$. In modern high-precision cosmology this is one of the most delicate issue when modeling of the observed universe.  
In this paper we provide a number of mathematical results that allow to compare the past lightcone regions $\mathcal{C}^-_L(p, {g})$ and $\mathcal{C}^-_L(p, \hat{g})$. In particular, we introduce a scale-dependent \textit{lightcone     comparison    functional} $E_{\widehat\Sigma\Sigma}[\phi_L]$ between the physical  and the FLRW reference celestial spheres $\Sigma_L\,\subset\,\mathcal{C}^-_L(p, {g}) $ and $\widehat\Sigma_L\, \subset\, \mathcal{C}^-_L(p, \hat{g})$  probed, at the given length scale $L$, on the respective lightcones regions  $\mathcal{C}^-_L(p, {g})$ and $\mathcal{C}^-_L(p, \hat{g})$. It is important to stress that the scale-dependent map $\phi_L$, is not an abstract map, but it actually relates the physical observations on $\Sigma_L$ with those, described with a FLRW bias, 
on $\widehat\Sigma_L$.  The functional $E_{\widehat\Sigma\Sigma}[\phi_L]$  is defined by a harmonic map type energy and has a number of remarkable properties. In particular it vanishes iff , at the given length-scale $L$, the corresponding lightcone surface sections $\Sigma_L$ and $\widehat\Sigma_L$ (which are topologically  2-spheres, as long as null-caustics are absent)  are isometric. Moreover, the $\inf$ of  $E_{\widehat\Sigma\Sigma}[\phi_L]$, over a suitable class of  extended maps $\phi_L$ (extension necessary in order to account also for the presence of lightcone caustics),  provides a scale-dependent distance functional, $d_L[\widehat\Sigma, \Sigma]$,  between the physical and the FLRW reference lightcones $\mathcal{C}^-_L(p, {g})$ and $\mathcal{C}^-_L(p, \hat{g})$. This distance   significantly extends the lightcone theorem proved in \cite{ChoquetLightCone}.  Moreover,  we show that in the  caustic-free region near the tip $p$  of $\mathcal{C}^-_L(p, {g})$ and $\mathcal{C}^-_L(p, \hat{g})$,  namely for $L$ small enough,  $d_L[\widehat\Sigma, \Sigma]$ is related to the spacetime scalar curvatures $R(g)$ and $R(\hat{g})$, in the interior of these lightcones, a relation that   may play an important role in  cosmological modeling.

\section{Cosmological observers and observational coordinates  along the Past Lightcones}
\label{sect2}

Throughout this paper  $(M, g)$ denotes a  cosmological spacetime  where $g$ is a Lorentzian metric  (of signature $(+,+,+,-)$), and where $M$ is  a smooth $4$-dimensional manifold which for our purposes we can assume diffeomorphic to $\mathbb{R}^4$.  In local coordinates $\{x^i\}_{i=1}^4$, we write $g=g_{ik}dx^i\otimes dx^k$, where the metric components $g_{ik}\,:=\,g(\partial_i, \partial_k)$
 in the coordinate basis $\{\partial_{i}:=\partial /\partial  x^i\}_{i=1}^4$, have the Lorentzian signature $(+,+,+,-)$, and the Einstein summation convention is in effect. We denote by $\nabla_{(g)}$  (or  $\nabla$ if there is no danger of confusion) the Levi--Civita connection of $g$, and let $\mathcal{R}m(g)=\mathcal{R}^{i}_{klm}\,\partial _i\otimes dx^k\otimes dx^l\otimes dx^m$, $\mathcal{R}ic(g)=\mathcal{R}_{ab}\,dx^a\otimes dx^b$ and $\mathcal{R}(g)$ be the corresponding Riemann,  Ricci and  scalar curvature operators, respectively.   We assume that $(M, g)$  is associated with the evolution of a universe which is (statistically) isotropic and homogeneous on sufficiently large scales, whereas local inhomogeneities dominate over smaller scales.   The mass--energy content in $(M, g)$ is phenomenologically described by an energy-momentum tensor $T$  the explicit expression of which is not needed in our analysis, we only assume that its  matter components  characterize a  Hubble flow that generates a family of preferred world-lines parametrized by proper time $\tau$
\begin{eqnarray}
\label{observers}
\gamma_s\,:\,\mathbb{R}_{>0}\,&\longrightarrow & \,(M, g)\\
\tau\,&\longmapsto&\,\gamma_s(\tau)\;,\nonumber
\end{eqnarray}
 and labeled by suitable comoving (Lagrangian) coordinates  $s$.
 We set $c\,=\,1$,  and denote by $\dot{\gamma}_s\,:=\,\frac{d\gamma_s(\tau)}{d\tau}$,\,with \,$g(\dot\gamma_s,  \dot\gamma_s)\,=\,-1$, the corresponding  $4$-velocity field.  For simplicity, we assume that the worldlines (\ref{observers}) are geodesics, \textit{i.e.} $\nabla_{\dot\gamma_s}\,\dot\gamma_s\,=\,0$. 
This is the spacetime within which we can frame the actual cosmological data gathered from our past lightcone observations. If we adopt the weak form of the cosmological principle, $(M, g, \gamma_s)$ can be identified with the phenomenological background spacetime or \textit{Phenomenological Background Solution (PBS)},  according to the notation introduced in \cite{KolbMarraMatarrese}. In the same vein, we define \textit{Phenomenological Observers}  the collection of observers $\{\gamma_s\}$ comoving with the Hubble flow. 

\subsection{The phenomenological lightcone metric}
Since in our  analysis we fix our attention on  a given  observer, we drop the subscript $s$ in (\ref{observers})), and  describe a finite portion of  the observer's world-line with the timelike geodesic segment  $\tau\,\longmapsto\,\gamma(\tau)$,\,$-\delta<\tau<\delta$, \, for some $\delta>0$, \,\,where  $p\,:=\,\gamma(\tau=0)$ is  the selected observational event. To set up the appropriate coordinates along $\gamma(\tau)$,  let $\left(T_p M,\,g_p,\,\{E_{(i)}\}\right)$ be the tangent space to $M$ at $p$ endowed with a $g$-orthonormal frame $\{E_{(i)}\}_{i=1,\ldots,4}$,\,\,$g_p\left(E_{(i)}, E_{(k)}\right)=\eta_{ik}$, where $\eta_{ik}$ is the Minkowski metric, and where $E_{(4)}\,:=\,\dot{\gamma}(\tau)|_{\tau=0}$. Notice that  by parallel transport, this basis can be propagated along $\gamma(\tau)$.   Let us introduce the set of past-directed null vectors and the set of  past-directed causal vectors in $(T_pM, g_p)$ according to
\begin{equation}
\label{MinkLightCone}
C^-\left(T_pM, g_p \right)\,:=\,\left\{X\,=\,\mathbb{X}^iE_{(i)}\,\not=\,0\,\in\,T_pM\,\,|\,\,\mathbb{X}^4+r=0 \right\}\;,
\end{equation}
\begin{equation}
\label{pastMinkLightCone}
\overline{C^-}\left(T_pM, g_p \right)\,:=\,\left\{X\,=\,\mathbb{X}^iE_{(i)}\,\not=\,0\,\in\,T_pM\,\,|\,\,\mathbb{X}^4+r\,\leq\,0 \right\}\;,
\end{equation}
where $r:=(\sum_{a=1}^3(\mathbb{X}^a)^2)^{1/2}$.  We use these sets of vectors in order to introduce observational coordinates in (a region of) the causal past of $p$, ${J}^-(p, g)$, by exploiting the exponential mapping based at $p$,
\begin{eqnarray}
\label{expmapdef}
\exp_p\,:\,\overline{C^-}\left(T_pM, g_p \right)\,&\longrightarrow&\;\;\;\;M\\
X\;\;\;\;\;\;&\longmapsto&\;\;\;\;exp_p\,(X)\,:=\,\lambda_X(1)\nonumber\\
\end{eqnarray}
where $\lambda_X\,:\,[0, \infty)\,\longrightarrow\,(M, g)$ is the past-directed causal geodesic emanating from the point $p$ with initial tangent vector $\dot{\gamma}_X(0)\,=\, X\in\overline{C^-}\left(T_pM, g_p \right)$.  If we assume that the metric is sufficiently regular\footnote{A Lipschitz condition  for the metric components suffices.}, then there is  a neighborhood $N_0(g)$ of  $0$ in $T_pM$ and a geodesically convex neighborhood of $p$, $U_p \subset\,(M, g)$, defined by all points $q\in M$ which are within the domain of injectivity of $exp_p$,  where we can  introduce geodesic normal coordinates $(X^i)$ according to
\begin{eqnarray}
\label{geodnormal}
X^i\,:=\,\mathbb{X}^i\,\circ \,\exp_p^{-1}\,:\,M\cap\,U_p\,&\longrightarrow&\,\mathbb{R}^4\\
q\,\,\,\,\,\,\,&\longmapsto&\,X^i(q)\,:=\, \mathbb{X}^i\left(\exp_p^{-1}(q)\right)\nonumber
\end{eqnarray}
where $\mathbb{X}^i\left(\exp_p^{-1}(q)\right)$ are the components, in the $g$-orthonormal frame $\{E_{(i)}\}$, of the vector $\exp_p^{-1}(q)\,\in\,T_pM$. In particular, if we consider the past lightcone $\mathcal{C}^-(p,g)$  with vertex  at $p$, then 
away from the past null cut locus of $p$, \textit{i.e.} away from the set of lightcone caustics, normal coordinates can be used to parametrize the past light cone region  $\mathcal{C}^-(p,g)\cap\,U_p$, 
\begin{eqnarray}
\label{exp1}
\exp_p\,:\,C^-\left(T_pM,\,g_p\right)\cap\,N_0(g)\,&\longrightarrow&\,\mathcal{C}^-(p,g)\,\cap\,U_p\\
X\,=\,\mathbb{X}^i E_{(i)}\,\,\,\,\,\,\,\,\,\,&\longmapsto&\,\exp_p(\mathbb{X}^iE_{(i)})\,=\,q\, \Rightarrow \,\{X^i(q)  \}\,\nonumber\;.
\end{eqnarray}
Similarly, by restricting $\exp_p$ to $\overline{C^-}\left(T_pM,\,g_p\right)\cap\,N_0(g)$ we can parametrize with normal coordinates the region ${J}^-(p,g)\,\cap\,U_p$ within the causal past  ${J}^-(p,g)$ of $p$. In particular,  we can foliate   ${J}^-(p,g)\,\cap\,U_p$ with the family  of past lightcones $\mathcal{C}^-(\gamma(\tau),g)$ associated with the events $\gamma(\tau)\cap\,U_p$, \,\,$-\delta\,<\,\tau\,\leq\,0$,    along the observer past-directed world line. We can specialize the normal coordinates so introduced by setting
\begin{equation} 
x^1\,:=\,r\,:=\,\sqrt{\sum_{a=1}^3(X^a)^2},\,\,\,\,
x^2\,:=\,\theta\left({X^a}/{r}\right),\,\,\,\,
x^3\,:=\,\phi\left({X^a}/{r}\right),\,\,\,\,
x^4\,:=\,\tau\,=\,X^4\,+\,r\,,
\end{equation}
where $\theta\left({X^a}/{r}\right)$,\,\, $\phi\left({X^a}/{r}\right)$,\, $a=1,2,3$, denote the standard angular coordinates of the direction $\left({X^a}/{r}\right)$ on the unit 2-sphere $\mathbb{S}^2$ in $T_pM$ and where, according to (\ref{MinkLightCone}), $x^4= 0$ corresponds to the light cone region  $\mathcal{C}^-(p,g)\cap\,U_p$. Notice that  at the vertex $p\,=\,\gamma(\tau=0)$,  the coordinate function $x^4$ is not differentiable (but it is continuous). 
\begin{rem} 
 Under the stated hypotheses, and as long as we stay away from the vertex $p$ and from its null cut locus,  we have that the past lightcone region $\mathcal{C}^-(p, g)\,\cap\,U_p\setminus \{p\}$
 is topologically foliated by the $r$-dependent family of   2--dimensional surfaces $\Sigma(p, r)$, the \textit{celestial spheres at scale $r$}, reached by the set of  past directed null geodesics  as the affine parameter $r$  varies, \textit{i.e.},
\begin{equation}
\Sigma(p, r)\,:=\,\left\{\left.\exp_p\left(r\, \underline{n}\right)\,\right|\,\, \underline{n}:=(\theta,\,\phi) \,\in\,\mathbb{S}^2\,\subset\,T_pM\right\}\;.
\end{equation}
\end{rem}
Each $\Sigma(p, r)$ is topologically a 2-sphere endowed with the $r$-dependent family of  two-dimensional Riemannian metrics 
\begin{equation}
h(r)\,:=\,\left.\left(\exp_p^*\,g|_{\Sigma(p, r)}\right)_{\alpha\beta}\,dx^\alpha dx^\beta\right|_{r}
\end{equation}
obtained by using the exponential map to pull back  to $\mathbb{S}^2\subset T_pM$ the 2-dimensional metric $g|_{\Sigma(p, r)}$ induced on $\Sigma(p, r)$ by the embedding  $\Sigma(p, r)\,\hookrightarrow \,(M, g)$.  
We normalize this metric  by imposing that the angular variables $x^\alpha=(\theta, \phi)$, in the limit $r\,\searrow\, 0$, reduce to the standard spherical coordinates on the unit 2-sphere 
$\mathbb{S}^2$, \textit{i.e.},
\begin{equation}
\label{limitr1}
\left.\lim_{r\searrow 0}\right|_{x^4=0}\,\frac{h_{\alpha\beta}(r)\,dx^\alpha dx^\beta}{r^2}\,
=\,d\Omega^2\,:=\,d\theta^2\,+\,\sin^2\theta\,d\phi^2\;.
\end{equation}
For a physical interpretation \cite{Ellis2},  it is convenient to parametrize $h(r)$ as a sky-mapping metric 
\begin{equation}
\label{etametric0}
h(r)\,=\,D^2(r)\,\left(d\Omega^2\,+\,\mathcal{L}_{\alpha\beta}(r)dx^\alpha dx^\beta \right)\;,  
\end{equation}
where $d\Omega^2$ is the unit radius round metric on $\mathbb{S}^2$ (see (\ref{limitr1})), and the coordinates $\{x^\alpha\}_{\alpha=2,3}$ provide 
the direction of observation (as seen at $p$) of the  astrophysical sources on the celestial sphere  $\Sigma(p, r)$.\, The function $D(r)$ is the  observer area distance defined by the relation $d\mu_{h(r)}  =  D^2(r)\,d\mu_{\mathbb{S}^2}$ where $d\mu_{h(r)}$ is the pulled-back (via $\exp_p$) area measure of $\left(\Sigma(p, r),\,g|_{\Sigma(p, r)}\right)$, (roughly speaking, $d\mu_{h(r)}$ can be interpreted \cite{Ellis2} as the cross-sectional area element at the source location as seen by the observer at $p$) and $d\mu_{\mathbb{S}^2}$ is the area element on the unit round sphere $\mathbb{S}^2\in\,T_pM$  
(\textit{i.e.}, the element of solid angle subtended by the source at the observer location $p$).  In the same vein, the symmetric tensor  field $\mathcal{L}_{\alpha\beta}(r)$, describing the distortion of the normalized metric $h(r)/D^2(r)$ with respect to the round metric $d\Omega^2$,  can be interpreted as the image distortion of  the sources on $\left(\Sigma(p, r), h(r)\right)$ as seen by the observer at $p$. This term, which in general is not trace-free, involves both  the gravitational lensing shear \cite{Ellis2} and the gravitational focusing of the light rays generating the local source image magnification.  By taking into account these remarks, we have the following characterization of the past lightcone metric in a neighborhood of the point $p$.
\begin{lemma}
In the geometrical coordinates introduced above, the null geodesics generators of $\mathcal{C}^-(p, g)\,\cap\,U_{p}$  have equation $x^4\,=\,0$, $x^\alpha\,=\, const.$,  and  their tangent vector is provided  by $\frac{\partial}{\partial x^1}$,\,\, with $(\exp^*g)\left(\frac{\partial}{\partial x^1},\,\frac{\partial}{\partial x^1}\right)\,=\,0$. Since $\frac{\partial}{\partial x^1}$ is past-directed we can introduce the  normalization 
\begin{equation}
 \lim_{r\searrow 0}\,(\exp^*g)\left(\frac{\partial}{\partial x^1},\,\dot\gamma\right)\,=\,1\,
\end{equation}
and  write the restriction of the  spacetime metric $g$ on  $\mathcal{C}^-(p, g)\,\cap\,U_{p}$ according to\footnote{A detailed and very informative analysis of geodesic coordinates along the past light cone is provided by \cite{Fleury1}.}   
\begin{equation}
\label{observative1}
\left.g\right|_{x^4=0}\,=\,g_{44}\,(dx^4)^2+2g_{14}dx^1dx^4+2g_{4\alpha}dx^4dx^\alpha+h_{\alpha\beta}dx^\alpha dx^\beta\;,
\end{equation}
where $\alpha, \beta=2,3$, and where the components $g_{ik}(x^i):=(\exp^*g)(\frac{\partial}{\partial x^i},\,\frac{\partial}{\partial x^k})$, and $h_{\alpha\beta}(x^i)$ are all evaluated for $x^4\,=\,0$. 
\end{lemma}
 As already stressed, the coordinates $\{x^i\}$ are singular at the vertex $\gamma(\tau=0)=p$ of the cone. A detailed analysis of  the  limit $r\,\searrow\, 0$, besides the standard assumptions we already made, is carried out in detail in the fundational paper  \cite{EllisPhRep} (see paragraph 3) and in \cite{Choquet1}, (see paragraphs 4.2.1-4.2.3-4.5, the results presented there are stated for the future lightcone, but they can be easily adapted to the past lightcone).  
\begin{rem}
Clearly the lightcone metric (\ref{observative1})  does not hold when caustics form, however our final result involving the characterization of a distance functional between lightcones naturally extends to the case  when caustics are present.
  \end{rem}
\subsection{The reference FLRW lightcone metric }
Along the physical metric $g$, we also introduce in $M$ the FLRW metric $\hat{g}$ and the family of global Friedmannian observers $\hat{\gamma}_s$  that, at the homogeneity scale,  we can  associate with the cosmological  data. This is the \textit{Global Background Solution (GBS)} according to \cite{KolbMarraMatarrese}. In full generality the  geodesics ${\tau}\longmapsto{\gamma}({\tau})$,  and $\hat{\tau}\longmapsto\hat{\gamma}(\hat{\tau})$,\,$-\delta<, \tau,\,\hat{\tau}<\delta$, associated with the corresponding Hubble flow in $(M, g, \gamma)$ and  $(M, \hat{g}, \hat{\gamma})$, will be distinct but, in line with the set up adopted here, we assume that they share a common observational event $p$.  We normalize the proper times $\tau$ and $\hat{\tau}$ along ${\gamma}({\tau})$ and $\hat{\gamma}(\hat{\tau})$  so that at $\tau\,=\,0\,=\,\hat{\tau}$ we have $\gamma(0)\,=\,p\,=\,\hat{\gamma}(0)$.  Hence, together with  the coordinates $\{x^i \}$  in $(M, g, \gamma_s)$, describing the observational metric (\ref{observative1}) on the past lightcone   $\mathcal{C}^-(p,g)\,\cap\,U_{p}$, we  introduce corresponding (normal) coordinates $\{Y^k \}$ in the reference  $(M, \hat{g}, \hat{\gamma})$. With an obvious adaptation of the analysis for $(M, g)$, carried out in previous subsection, let
$N_0(\hat{g})$ denote the domain of injectivity  of  the exponential mapping  $\widehat{\exp}_{p}\,:\,T_{p}M\,\longrightarrow\,(M, \hat{g})$  based at the event $p=\hat{\gamma}(0)$. If  $\hat{U}_{p}\subset\,(M, \hat{g})$ denotes  the region of injectivity of  $\widehat{\exp}_{p}$ we can consider normal coordinates
\begin{equation}
\label{geodnormalY}
Y^i\,:=\,\mathbb{Y}^i\,\circ \,\widehat{\exp}_p^{-1}\,:\,(M, \hat{g})\cap\,\hat{U}_{p}\,\longrightarrow\,\mathbb{R}\,,
\end{equation}
where  $\mathbb{Y}^i$ are the components of  the vectors $\mathbb{Y}\in\,T_pM$ with respect to a $\hat{g}$-orthonormal frame $\{\hat{E}_{(i)}\}_{i=1,\ldots,4}$ with $\hat{E}_{(4)}\,:=\,\hat{\dot{\gamma}}(0)$. Within $\hat{U}_{p}$  we can introduce, in full analogy with (\ref{exp1}) and  (\ref{observative1}),   the coordinates   $y^1\,:=\,\hat{r}\,=\,(\sum_{a=1}^3(Y^a)^2)^{1/2}$,  
\,$y^\alpha|_{\alpha=2,3}\,=\,\left(\theta\left({Y^a}/\hat{r}\right),\,\phi\left({Y^a}/\hat{r}\right)\right)$ and parametrize
 $\mathcal{C}^-(p,\hat{g})\,\cap\,\hat{U}_{p}$ in terms of the 2-dimensional spheres 
 \begin{equation}
\hat{\Sigma}(p, \hat{r})\,:=\,\left\{\left.\hat{\exp_p}\left(\hat{r}\, \underline{n}\right)\,\right|\,\, \underline{n}:=(\theta,\,\phi) \,\in\,\mathbb{S}^2\,\subset\,T_pM\right\}\;,
\end{equation}
 endowed with the round metric
 \begin{equation}
 \label{etametric01}
\hat{h}(\hat{r})\,:=\,\left. (\hat{g})_{\alpha\beta}dy^\alpha dy^\beta\right|_{\hat{r}}\,=\,
a^2(\hat{r})\,\hat{r}^2\left(d\theta^2\,+\,\sin^2\theta d\phi^2\right)\;,
\end{equation} 
where $a(\hat{r})$ is the  FLRW expansion factor corresponding to the distance $\hat{r}$. Hence,
we can  write the  metric $\hat{g}$ on the reference FLRW past lightcone region $\mathcal{C}^-(p,\hat{g})\,\cap\,\hat{U}_{p}$  as
\begin{equation}
\label{observative2}
\left.\hat{g}\right|_{y^4=0}\,=\,\hat{g}_{44}\,(dy^4)^2+\hat{h}_{\alpha\beta}dy^\alpha dy^\beta\;.
\end{equation}

\section{Comparing lightcones: a scale dependent comparison functional}
According to our hypotheses,  the spacetime $(M, g, \gamma_s)$  describes the evolution of a universe which is isotropic and homogeneous only at sufficiently large scales $L_0$. At these homogeneity scales  $(M, g, \gamma_s)$ is modeled by  the FLRW spacetime $(M, \hat{g}, \hat{\gamma}_s)$. Even if at smaller scales, where inhomogeneities statistically dominate,  $(M, g, \gamma_s)$ provides the \textit{bona fide} spacetime describing cosmological observations,  we can still use the reference $(M, \hat{g}, \hat{\gamma}_s)$ as a background  FLRW model.  As the observational length scale $L$ varies  from the local highly inhomogeneous  regions to the homogeneity scale $L_0$, we do not assume a priori that  $(M, g, \gamma_s)$ is perturbatively near to the reference  FLRW  spacetime $(M, \hat{g}, \hat{\gamma}_s)$. Rather, we compare $(M, g, \gamma_s)$ with $(M, \hat{g}, \hat{\gamma}_s)$ keeping track of the pointwise and global relations among the various geometric quantities involved. In particular,  we will compare  the lightcone region $\mathcal{C}^-(p, {g})\,\cap\,{U}_{p}$  with the reference FLRW lightcone  region   $\mathcal{C}^-(p,\hat{g})\,\cap\,\hat{U}_{p}$,  assuming that in such a range there are no lightcone 
caustics.  As already emphasized, this is an assumption that makes easier to illustrate some of the technical arguments presented here, in the final part of the paper we indicate how our main result, concerning the existence and the properties of the distance functional described in the introduction, holds also in the more general case when caustics are present.  That said,  let us consider the following scale--dependent subsets of  the past light cones $\mathcal{C}^-(p,\,g)$ and $\mathcal{C}^-(p,\, \hat{g})$,
\begin{equation}
\label{ConesRegions}
\mathcal{C}^-_{L}(p,g)\,:=\,\exp_p\left[C^-_{L\leq\,L_0}\left(T_pM, g_p \right) \right],\,\,\,
\mathcal{C}^-_{L}(p,\hat{g})\, :=\,\widehat{\exp}_p\left[C^-_{L\leq\,L_0}\left(T_pM, \hat{g}_p \right)\right]\,,
\end{equation}
 where 
\begin{eqnarray}
 C^-_{L\leq\,L_0}\left(T_pM, g_p \right)\,&:=&\,  \left\{X\,=\,\mathbb{X}^i E_{(i)}\,\in\,(T_pM, g_p)\,|\,X^4\,+\,r\,=\,0,\;-L_0\,\leq\,X^4\,\leq\,0  \right\}\,,\\
 \nonumber\\
 C^-_{L\leq\,L_0}\left(T_pM, \hat{g}_p \right)\,&:=&\,  \left\{Y\,=\,\mathbb{Y}^a \hat{E}_{(a)}\,\in\,(T_pM, \hat{g}_p)\,|\,Y^4\,+\,\hat{r}\,=\,0,\;-L_0\,\leq\,Y^4\,\leq\,0  \right\}\,,
\end{eqnarray}
are the exponential map domains associated with the observational length-scale $L$ up to the homogeneity scale $L_0$.  Under the stated caustic--free assumption, both $\mathcal{C}^-_{L}(p,{g})$ and $\mathcal{C}^-_{L}(p,\hat{g})$  can be foliated  in terms of the 2-dimensional surfaces 
${\Sigma}(p, {r})$ and $\hat{\Sigma}(p, \hat{r})$ introduced in the previous section,  \textit{i.e.}, we can write
$$
\mathcal{C}^-_{L}(p,g)\,=\,\bigcup_{\,0\,\leq\,r\,\leq\,L_0}\,{\Sigma}(p, {r}),\,\,\,\,\,
\mathcal{C}^-_{L}(p,\hat{g})\,=\,\bigcup_{\,0\,\leq\,\hat{r}\,\leq\,L_0}\,\hat{\Sigma}(p, \hat{r})\;.
$$
On $\mathcal{C}^-_{L}(p,g)$ and  $\mathcal{C}^-_{L}(p,\hat{g})$ the normal coordinates\footnote{We use the letters from the first half of the alphabet, $a,b,c,d, \ldots$ to index the coordinates $\{y\}$; the letters from the second half $i, j, k, \ell, \ldots$ provide the indexing of the  coordinates $\{x\}$.} $\{x^i \}$  and   $\{y^a \}$,  associated with the observational metric (\ref{observative1})  and the reference metric (\ref{observative2}),  cannot be directly identified since they
 are defined in terms of the distinct exponential mappings $\exp_p$ and $\widehat{\exp}_p$ and, for a given initial tangent vector \, 
$X\,\in\,C^-_{L\leq\,L_0}\left(T_pM, {g}_p \right)\cap\,C^-_{L\leq\,L_0}\left(T_pM, \hat{g}_p \right)$, we have 
\begin{equation}
\exp_p(X)\,=\,q\,\not=\,\widehat{\exp}_p(X)\,=\,\hat{q}\;.
\end{equation}
However, $q$ and $\hat{q}$ are in the open spacetime region defined by 
\begin{equation}
 M_p\,:=\,\exp_p\left(N_0({g}) \right)\,\cap\,\hat{\exp_p}\left(N_0(\hat{g}) \right)\subset\,M\;,
\end{equation}
and since  $\exp_p$ and $\widehat{\exp}_p$ are local diffeomorphisms from $N_0({g})\cap\,N_0(\hat{g})\subset\,T_pM$ into $M_p$,  the map defined by
\begin{eqnarray}
\label{psi0}
\psi\,:(M_p\,\cap\,\mathcal{C}^-_{L}(p, \hat{g}),\, \hat{g})\,&\longrightarrow&\,(M_p\,\cap\,\mathcal{C}^-_{L}(p,{g}),\, {g})\\
\hat{q}\,&\longmapsto&\,\psi(\hat{q})\,=\,{q}\,=\,{\exp}_p\left(\widehat{\exp}_p^{- 1}(\hat{q}) \right)\nonumber
\end{eqnarray}
 is a diffeomorphism with $\psi(p)\,=\,\mathrm{id}_M$. In particular, in terms of the coordinates  $\{x^i \}$ and $\{y^a \}$ we can locally write  
\begin{equation}
\label{psi1}
y^a(\hat{q})\,\longmapsto\,x^i({q})\,=\,\psi^i(y^b(\hat{q})).
\end{equation}
In order to describe  at a given length scale $0\,<\,L\,\leq\,L_0$, the effect of these diffeomorphisms on the lightcone regions   $\mathcal{C}^-_{L}(p,g)$ and $\mathcal{C}^-_{L}(p,\hat{g})$,  let us consider the spherical surfaces
\begin{equation}
\label{celestialspheres}
\left(\Sigma_L,\,h\right)\,:=\,[\Sigma(p, {r=L}),\, h],\,\,\,\,\,(\widehat\Sigma_L,\,\hat{h})\,:=\, [\widehat{\Sigma}(p, \hat{r}=L),\,\hat{h}]
\end{equation}
 with their respective metrics $h$ and $\hat{h}$, and where, since the notation wants to travel light, we drop the explicit reference to the vertex $p$ of the lightcone and  where we have replaced the affine parameters $r$ and $\hat{r}$ with the preassigned value $L$ of the probed length scale. The surfaces $\left(\Sigma_L,\,h\right)$ 
and $(\widehat\Sigma_L,\,\hat{h})$  characterize, at the given scale $L$,  the celestial sphere at $p$  as seen  by the physical observer  and by the reference FLRW observer, respectively.\\

A direct application of the standard geometrical set-up of harmonic map theory  (see \emph{e.g.} \cite{jost})  provides the following notational lemma directly connecting our analysis to harmonic maps between surfaces.
\begin{lemma}
Let $\psi_L$ be the diffeomorphism $\psi$ restricted to the surfaces  $(\widehat{\Sigma}_L,\,\hat{h})$  and $(\Sigma_L,\, h)$,
\begin{equation}
\label{Map1}
\psi_L\,:\,(\widehat{\Sigma}_L,\,\hat{h})\,\longrightarrow\,(\Sigma_L,\, h)
\end{equation}
then  we can introduce the  pull--back  bundle $\psi_L^{-1}T\widehat{\Sigma}_L$ whose sections $v\equiv \psi_L^{-1}V:= V\circ \psi_L$,\,  $V\in C^{\infty }(\widehat\Sigma, T\widehat\Sigma_L)$,\, are the vector fields over $\widehat\Sigma$ covering the map $\psi_L$. If $T^*\widehat\Sigma_L$ denotes the cotangent bundle to $(\widehat\Sigma_L, \hat{h})$, then the differential $d\psi_L\,=\,\frac{\partial\psi^i_L}{\partial y^{a}}dy^{a}\otimes \frac{\partial }{\partial \psi^i}$ can  be interpreted as a section of  $T^*\widehat\Sigma_L\otimes \psi^{-1}T\Sigma_L$, and its
Hilbert--Schmidt norm, in the bundle metric
\begin{equation}
\label{bundmetr}
\langle\cdot ,\cdot \rangle_{T^*\widehat\Sigma_L\otimes \psi^{-1}T\Sigma_L}\,:=\,
\hat{h}^{-1}(y)\otimes h(\psi_L(y))(\cdot ,\cdot )\;,
\end{equation}
is provided by  
\begin{equation}
\label{HSnorm}
\langle d\psi_L , d\psi_L \rangle_{T^*\widehat\Sigma_L\otimes \psi^{-1}T\Sigma_L}\,=\,
\hat{h}^{ab}(x)\,\frac{\partial \psi^{i}(y)}{\partial y^{a}}
\frac{\partial\psi^{j}(y)}{\partial y^{b}}\,h_{ij}(\psi(y))=\,tr_{\hat{h}(y)}\,(\psi_L^{*}\,h)\;,
\end{equation} 
where
\begin{equation}
\label{pullcomp}
\psi_L^* h\,\Longrightarrow\,
\left(\psi_L^* h\right)_{ab}\,=\,\frac{\partial\psi^i(y^c)}{\partial y^a}\frac{\partial\psi^k(y^d)}{\partial y^b}\,h_{ik}
\end{equation}
provides the pull-back  of the metric ${h}$ on  $\widehat{\Sigma}_L$. 
\end{lemma}
The connection between  the pulled-back metric $\psi_L^* h$  and the round metric $\hat{h}$, both defined on  $\widehat{\Sigma}_L$, is provided by the following proposition where we respectively denote by $R_L(\hat{h})$ and  $R_L( h)$  the scalar curvature of  $(\widehat\Sigma_L, \hat{h})$ and $(\Sigma_L, h)$, and we let $\Delta_{\hat{h}}\,:=\,\hat{h}^{\alpha\beta}\nabla_\alpha\nabla_\beta$ be the Laplace-Beltrami operator on $(\widehat\Sigma_L, \hat{h})$.  Notice that the scalar curvature $R_L(\hat{h})$ is associated with the metric (\ref{etametric01}) evaluated for $\hat{r}\,=\,L$ and hence is given by the constant $R_L(\hat{h})\,=\,\frac{2}{a^2(L)\,L^2}$. In a similar way, $R_L( h)$ is associated with the metric (\ref{etametric0}) evaluated for 
$r\,=\,L$, and as such it depends on the area distance $D^2(L)$ and on the lensing distortion $\mathcal{L}_{\alpha\beta}(L)dx^\alpha dx^\beta$.
\begin{prop}
\label{3points}
Let ${q}_{(i)}$,\,$i\,=1,2,3$ three distinct points intercepted, on the observer celestial sphere $({\Sigma}_L,\,{h})$, by three past-directed null geodesics on $\mathcal{C}^-_{L}(p,{g})$, and
let $\hat{q}_{(i)}$,\,$i\,=1,2,3$, three distinguished points on the reference FLRW celestial sphere  
$(\widehat{\Sigma}_L,\,\hat{h})$, characterizing three corresponding past-directed null directions on $\mathcal{C}^-_{L}(p,\hat{g})$.  If
$\zeta\,\in\,PSL(2, \mathbb{C})$ denotes the fractional linear transformation in the projective special linear group, describing the automorphism of  $(\widehat{\Sigma}_L,\,\hat{h})$ that brings  $\{\psi^{-1}(q_i)\}$ into $\hat{q}_{(i)}$,\,$i\,=1,2,3$,  then there is a positive scalar function ${\Phi}_{\widehat{\Sigma}\Sigma}\,\in\,C^\infty(\hat{\Sigma},\,\mathbb{R})$,
solution of the elliptic partial differential equation
\begin{equation}
\label{lapconf}
-\,\Delta _{\hat{h}}\ln({\Phi}_{\widehat{\Sigma}\Sigma}
^2)\,+\,R_L(\hat{h})\,=\, R_L(h)\,{\Phi}_{\widehat{\Sigma}\Sigma}
^2\;,
\end{equation}
such that $\psi_L\circ \zeta$ characterizes a conformal diffeomorphism between $(\widehat{\Sigma}_L,\,\hat{h})$  and $(\Sigma_L,\, h)$, \textit{i.e.}  
\begin{equation}
\label{psiconf0}
\left(\psi_L\circ \zeta\right)^* h\,=\,{\Phi}_{\widehat{\Sigma}\Sigma}
^2\,\hat{h}\;.
\end{equation}
\end{prop}
\begin{proof}
This is a direct consequence of  the   Poincare-Koebe uniformization theorem which implies that the 2-sphere with the pulled back metric $(\widehat{\Sigma}_L,\,\psi_L^* h)$   can be mapped conformally, in a one-to-one way, onto the round 2-sphere $(\widehat{\Sigma}_L,\,\hat{h})$. Recall that on the unit sphere $\mathbb{S}^2$, with its canonical round metric $d\Omega^2$, there is a unique conformal class $[d\Omega^2]$ and that the metric (\ref{etametric01}) on $\widehat{\Sigma}_L\simeq \mathbb{S}^2$,  rescaled according to  $\hat{h}/(a^2(\hat{r})\,\hat{r}^2)$,   is isometric to $d\Omega^2$. Hence, by the uniformization theorem, all metrics on  $\widehat{\Sigma}_L\simeq \mathbb{S}^2$ may be pulled back by diffeomorphisms to the conformal class $[\hat{h}]$  of the round metric with the chosen radius $a^2(\hat{r})\,\hat{r}^2$. Since $(\widehat{\Sigma}_L,\, \hat{h}/(a^2(\hat{r})\,\hat{r}^2))\simeq \mathbb{S}^2$,  the pullback is unique modulo the action of the conformal group group of the sphere $\mathrm{Conf}(\mathbb{S}^2)$. If we denote by $\mathcal{P}_{\mathbb{S}^2}$ the stereographic projection (from the north pole $(0,0,1)$ of  $\mathbb{S}^2:=\{(x,y,z)\in\mathbb{R}^3\,\,|\,\,x^2+y^2+z^2=1\}$)
\begin{equation}
\mathcal{P}_{\mathbb{S}^2}\,:\,\mathbb{S}^2\,\subset\,\mathbb{R}^3\,\longrightarrow\,\mathbb{C}\cup\{\infty \},\,\,\,\,
\mathcal{P}_{\mathbb{S}^2}(x,y,z)\,=\,\frac{x+i\,y}{1-z}\,,
\end{equation}
then we can identify $\mathrm{Conf}(\mathbb{S}^2)$ with the 6-dimensional projective special linear group $PSL(2, \mathbb{C})$ describing the automorphisms of    $\mathbb{S}^2\simeq\mathbb{C}\cup\{\infty \}$.    The elements of  $PSL(2, \mathbb{C})$ are the fractional linear transformations the Riemann sphere $\mathbb{S}^2\,\simeq\,\mathbb{C}\,\cup\,\{\infty\}$
\begin{eqnarray}
\mathbb{C}\,\cup\,\{\infty\}\,&\longrightarrow&\,\mathbb{C}\,\cup\,\{\infty\}\\
z\,&\longmapsto&\,\zeta (z)\,:=\,\frac{az+b}{cz+d}\,,\,\,\,\,\,a, b, c, d\,\in\,\mathbb{C}\,,\,\,\,ad\,-\,bc\,\not=\,0\,.\nonumber
\end{eqnarray}
These transformations  act on the diffeomorphism  (\ref{Map1}) according to
\begin{eqnarray}
\label{psl2caction}
PSL(2, \mathbb{C})\times(\widehat{\Sigma}_L,\,\hat{h})\,&\longrightarrow&\,(\Sigma_L,\, h)\\
\left( \zeta,\,y  \right)\,&\longmapsto&\,\psi_L(\zeta(y))\nonumber
\end{eqnarray}
where, abusing notation, we have denoted by $\zeta(y)$ the action that the fractional linear transformation $\zeta(z)$ defines on the point $y\in\widehat\Sigma_L$ corresponding, via stereographic projection, to the point $z\in\mathbb{C}\,\cup\,\{\infty\}$. 
This action may be a potential source of a delicate problem since  $PSL(2,\mathbb{C})$ is non-compact and ${\Phi}_{\widehat\Sigma\Sigma}$ is evaluated on the 
composition $\psi_L\circ \zeta$  defined by (\ref{psl2caction}). This is not problematic as long as $\zeta$ varies in the maximal compact subgroup of $PSL(2,\mathbb{C})$ generated by the isometries of  $(\widehat\Sigma, \hat{h})$. However, if we consider a sequence   $\{\zeta_k\}_{k\in\,\mathbb{N}}\in PSL(2,\mathbb{C})$ defined by larger and larger dilation (corresponding to larger and larger (local) Lorentz boosts of the surface $\widehat\Sigma$ in the reference spacetime $(M, \hat{g})$), then  the composition $\psi_L\circ \zeta_k$  may generate a sequence of conformal factors $\{{\Phi^2_{(k)}}_{\widehat{\Sigma}\Sigma}\}$ converging to a non-smooth function. To avoid these pathologies we exploit the fact that a linear fractional transformation is fully determined if we fix its action on three distinct points of the sphere. In our setting this corresponds to fixing the action on three distinct null direction in the lightcone region  $\mathcal{C}^-_{L}(p,\hat{g})$. In physical terms  this is equivalent to require that the FLRW reference observer at $p$ has to adjust his velocity and orientation in such a way that  three given astrophysical sources of choice  are in three specified position on the celestial sphere $(\widehat\Sigma_L, \hat{h})$ at scale $L$.  This is a gauge fixing of the action of  $PSL(2,\mathbb{C})$ that corresponds in a very natural way to adjust the location of three  reference observations  in order  to be able to compare the data on the physical past lightcone   
$\mathcal{C}^-_{L}(p,{g})$  with the data on the reference past lightcone $\mathcal{C}^-_{L}(p,\hat{g})$. 
 By fixing in this way the $PSL(2,\mathbb{C})$ action, the pullback  $\left(\psi_L\circ \zeta\right)^* h$  on $\hat{\Sigma}_L$ of the metric $h$ is well defined. By the Poincare-Koebe uniformization theorem the metric $\left(\psi_L\circ \zeta\right)^* h$ is in the same conformal class of $\hat{h}$. Let us denote by ${\Phi}^2_{\widehat{\Sigma}\Sigma}\,\in\,C^\infty(\hat{\Sigma},\,\mathbb{R})$ the corresponding conformal factor such that 
 $\left(\psi_L\circ \zeta\right)^* h\,=\,{\Phi}^2_{\widehat{\Sigma}\Sigma}\,\hat{h}$. If we set $e^f\,:=\,{\Phi}^2_{\widehat{\Sigma}\Sigma}\,$, then the  properties of the scalar curvature under the conformal transformation $h\,=\,e^f\,\hat{h}$ (see \textit{e.g.}, \cite{Berger})  provide the relation
 \begin{equation}
 \label{conftransf}
 R\left(\left(\psi_L\circ \zeta\right)^* h\right)\,=\,e^{\,-f}\,\left[ R(\hat{h})\,+\,\Delta_{\hat{h}}\,f \right]\;.
 \end{equation}
If for notational ease we keep on writing $R(h)$ for $R\left(\left(\psi_L\circ \zeta\right)^* h\right)=R(h(\psi_L\circ \zeta))$, then it follows from 
(\ref{conftransf}) that ${\Phi}^2_{\widehat{\Sigma}\Sigma}$ necessarily is a solution on $(\widehat\Sigma_L, \hat{h})$ of 
the elliptic partial differential equation  (\ref{lapconf}), solution that under the stated hypotheses always exists \cite{Berger}.
\end{proof}
According to the above result, there is a positive scalar function ${\Phi}_{\widehat{\Sigma}\Sigma}\,\in\,C^\infty(\hat{\Sigma},\,\mathbb{R})$ such that $\psi_L\circ \zeta$ characterizes a conformal diffeomorphism between $(\hat{\Sigma}_L,\,\hat{h})$ and $(\Sigma_L,\, h)$.  In components (\ref{psiconf0}) can be  written as    
\begin{equation}
\label{psiconf}
\left(\left(\psi_L\circ \zeta\right)^* h\right)_{ab}\,=\,\frac{\partial\psi_L^i(\zeta(y))}{\partial y^a}\frac{\partial\psi_L^k(\zeta(y))}{\partial y^b}\,h_{ik}\,=\,{\Phi}_{\widehat{\Sigma}\Sigma}^2\,\hat{h}_{ab}\;.
\end{equation}
 It follows that by tracing (\ref{psiconf}) with respect to $\hat{h}^{ab}$, we can express ${\Phi}_{\widehat{\Sigma}\Sigma}^2$ in terms of the Hilbert--Schmidt norm of
the differential $d\left(\psi_L\circ \zeta\right)\,=\,\frac{\partial\psi_L^i(\zeta(y))}{\partial y^{a}}dy^{a}\otimes \frac{\partial }{\partial \psi_L^i}$ according to  (see (\ref{HSnorm}))
 \begin{equation}
\label{Fdensity}
{\Phi}_{\widehat{\Sigma}\Sigma}^2\,=\,tr_{\hat{h}(y)}\,\left(\left(\psi_L\circ \zeta\right)^* h\right)\,=\,
\frac{1}{2}\,\hat{h}^{ab}\,\frac{\partial\psi_L^i(\zeta(y))}{\partial y^a}\frac{\partial\psi_L^k(\zeta(y))}{\partial y^b}\,h_{ik}\;.
\end{equation}
From  (\ref{psiconf}) we get $\det \left(\left(\psi_L\circ \zeta\right)^* h\right)\,=\, {\Phi}_{\widehat{\Sigma}\Sigma}^4\,\det (\hat{h})$, hence we can equivalently write the conformal factor as the  Radon-Nikodym derivative of the riemannian measure $d\mu_{\psi^*{h}}:=\left(\psi_L\circ \zeta\right)^*d\mu$, of  $(\hat{\Sigma}, \left(\psi_L\circ \zeta\right)^*h)$,  with respect to the riemannian measure  $d\mu_{\hat{h}}$ of the round metric $(\hat\Sigma, \hat{h})$,  \textit{i.e.}, 
\begin{equation}
\label{RadonNphi}
{\Phi}_{\widehat{\Sigma}\Sigma}^2\,=\,\frac{d\mu_{\psi^*{h}}}{d\mu_{\hat{h}}}\,=\,\frac{\left(\psi_L\circ \zeta\right)^*d\mu_h}{d\mu_{\hat{h}}}\;.
\end{equation}
Equivalently, this states that ${\Phi}_{\widehat{\Sigma}\Sigma}^2$ can be interpreted as the Jacobian of the  map $\psi_L\circ \zeta$,
\begin{equation}
\label{Jac}
{\Phi}_{\widehat{\Sigma}\Sigma}^2\,=\,\mathrm{Jac}(\psi_L\circ \zeta)\,.
\end{equation}
Along the same lines, we can associate to the inverse diffeomorphism
\begin{eqnarray}
\label{invpsi}
\left(\psi_L\circ \zeta\right)^{-1}\,:\,(\Sigma_L,\, h) \,&\longrightarrow&\, (\widehat{\Sigma}_L,\,\hat{h})\\
x\,&\longmapsto&\,\zeta^{-1}\left(\psi_L^{-1}(x) \right)\nonumber
\end{eqnarray}
a positive scalar function $\Phi_{\Sigma\widehat\Sigma}\,\in\,C^\infty({\Sigma},\,\mathbb{R})$ such that we can write
\begin{equation}
\label{confphiSigma}
\left(\left(\psi_L\circ \zeta\right)^{-1}\right)^* \hat{h}\,=\,{\Phi}_{\Sigma\widehat\Sigma}^2\,{h}\,,
\end{equation}
with
\begin{equation}
\label{gradients}
{\Phi}_{\Sigma\widehat\Sigma}^2\,=\,\frac{1}{2}\,{h}^{ik}\,\frac{\partial\left(\zeta^{-1}\left(\psi_L^{-1}(x) \right) \right)^a}{\partial x^i}
\frac{\partial\left(\zeta^{-1}\left(\psi_L^{-1}(x) \right) \right)^b}{\partial x^k}\,\hat{h}_{ab}\,=\, \frac{d\mu_{(\psi^{-1})^*{\hat{h}}}}{d\mu_{h}}\,=\,\frac{\left(\left(\psi_L\circ \zeta\right)^{-1}\right)^*d\mu_{\hat{h}}}{d\mu_h}\;.
\end{equation}
\\
To measure the global deviation of the conformal diffeomorphisms ${\Phi}_{\widehat\Sigma\Sigma}$ from an 
isometry between $(\hat{\Sigma}_L,\,\hat{h})$ and $(\Sigma_L,\, h)$ we  introduce  the following       comparison    functional where, for later use, we keep track of  the  $\zeta\,\in\,PSL(2, \mathbb{C})$ dependence in  ${\Phi}_{\widehat\Sigma\Sigma}$.
\begin{defn} (The lightcone       comparison    functional at scale $L$) \\ 
\label{sepfunct}
Let ${\Phi}_{\widehat{\Sigma}\Sigma}\,\in\,C^\infty(\hat{\Sigma},\,\mathbb{R})$ (or at least $C^2(\hat{\Sigma},\,\mathbb{R})$) be the positive scalar function  such that $\psi_L\circ \zeta$ characterizes the conformal diffeomorphism $\left(\psi_L\circ \zeta\right)^* h\,=\,{\Phi}_{\widehat{\Sigma}\Sigma}
^2\,\hat{h}$ between $(\widehat{\Sigma}_L,\,\hat{h})$  and $(\Sigma_L,\, h)$, then the associated lightcone      comparison    functional at scale $L$ is defined by
 \begin{equation}
\label{enfunct1}
E_{\widehat\Sigma\Sigma}[\psi_L,\,\zeta]\,:=\,\int_{\widehat{\Sigma}_L}({\Phi}_{\widehat\Sigma\Sigma}\,-\,1 )^2\,d\mu_{\hat{h}}\;.
\end{equation} 
\end{defn}
The functional  $E_{\widehat\Sigma\Sigma}[\psi_L,\,\zeta]$ is clearly related to the familiar harmonic map energy associated with the map 
$\psi_L\circ \zeta\,:\,\hat{\Sigma}\,\longrightarrow\,\Sigma$.  Explicitly, if  we take into account 
 (\ref{Fdensity}) we can write
\begin{equation}
\label{HamEnergy}
\int_{\widehat{\Sigma}_L} \widehat{\Phi}_L^2\,d\mu_{\hat{h}}\,=\,\frac{1}{2}\,\int_{\widehat{\Sigma}_L}
\hat{h}^{ab}\,\frac{\partial\psi_L^i(\zeta(y))}{\partial y^a}\frac{\partial\psi_L^k(\zeta(y))}{\partial y^b}\,h_{ik}\,d\mu_{\hat{h}}\;,
\end{equation}
which provides the harmonic map functional whose critical point  are the harmonic maps of the Riemann surface $(\hat{\Sigma}_L,\,[\hat{h}])$ into $(\Sigma_L,\, h)$, where $[\hat{h}]$ denotes the conformal class of the metric $\hat{h}$. Notice that,  whereas the harmonic map energy (\ref{HamEnergy}) is a conformal invariant quantity,  the functional $E_{\widehat\Sigma\Sigma}[\psi_L,\,\zeta]$ is not conformally invariant. Under a conformal trasformation $\hat{h}\,\longrightarrow\, e^{2f}\,\hat{h}$ we get
\begin{equation}
\int_{\widehat{\Sigma}_L}\left(e^{\,-\,f}{\Phi}_{\widehat\Sigma\Sigma}\,-\,1 \right)^2\,e^{2f}\,d\mu_{\hat{h}}\;.
\end{equation}
It is also clear from its definition that corresponding to large gradients  (see (\ref{gradients})),  $E_{\widehat\Sigma\Sigma}[\psi_L,\,\zeta]$ tends to the harmonic map energy. In this connection, it is important to stress that rather than on the space of smooth maps $C^\infty(\widehat\Sigma,\,\Sigma)$, the functional  $E_{\widehat\Sigma\Sigma}[\psi_L,\,\zeta]$ is naturally defined  on the
Sobolev space of maps $W^{1, 2}(\widehat\Sigma,\,\Sigma)$ which  are, together with their weak derivatives, square integrable.
This characterization, familiar when studying weakly-harmonic maps  \cite{helein} and which we discuss in detail below when minimizing  $E_{\widehat\Sigma\Sigma}[\psi_L,\,\zeta]$,  is important in our case when extending our analysis 
to the low regularity setting when lightcone caustics are present.\\
\begin{rem}
 It must be stressed that energy functionals such 
as (\ref{enfunct1})  are rather familiar in the problem of comparing shapes of surfaces in relation with computer graphic and visualization problems (see \textit{e.g.} \cite{JinYauGu} and \cite{YauGu}, to quote two  relevant papers in a vast literature). In particular,  (\ref{enfunct1})  has been introduced under the name of \textit{elastic energy} in an inspiring paper by J. Hass and P. Koehl  \cite{HassKoehl}, who use it as a building block of a more complex functional relevant to surface visualization. \\
\end{rem}
In our particular framework, the functional $E_{\widehat\Sigma\Sigma}[\psi_L,\,\zeta]$ has a number of important properties that make  it a natural candidate for comparing, at the given length scale $L$, the physical lightcone region $\mathcal{C}^-_{L}(p,{g})$  with the FLRW reference region $\mathcal{C}^-_{L}(p,\hat{g})$.  To start with, we prove the following general properties (in the smooth setting) . 
\begin{lemma}
\label{propfunct}
The functional  $E_{\widehat\Sigma\Sigma}[\psi_L,\,\zeta]$ is symmetric
\begin{equation}
\label{symmetryE}
E_{\widehat\Sigma\Sigma}[\psi_L,\,\zeta]\,=\,E_{\Sigma\widehat\Sigma}[\psi^{-1}_L,\,\zeta^{-1}]\;,
\end{equation}
where
\begin{equation}
\label{invenfunct1}
E_{\Sigma\widehat\Sigma}[\psi^{-1}_L,\,\zeta^{-1}]\,:=\,\int_{{\Sigma}_L}({\Phi}_{\Sigma\widehat\Sigma}\,-\,1 )^2\,d\mu_{{h}}\;,
\end{equation}
 is the      comparison    functional associated with the inverse map $(\psi_L\circ \zeta)^{-1}\,:\,{\Sigma}_L\,\longrightarrow\,\hat\Sigma_L$.\\
If $(\widetilde{\Sigma}_L,\, \tilde{h})$  is a third surface on the past 
lightcone  $\widetilde{\mathcal{C}}^-_{L_0}(p,\tilde{g})$, with vertex at $p$, associated with yet another reference FLRW metric $\tilde{g}$ on $M$ (say another member of the FLRW family of spacetimes, distinct from $\hat{g}$), and $\sigma_L\,:\Sigma_L\,\longmapsto\,\widetilde\Sigma_L$, \,$\Phi_{\Sigma\widetilde\Sigma}$ respectively are the corresponding diffeomorphism and conformal factor, then to the composition of maps
\begin{equation}
\widehat{\Sigma}_L\,\underset{\psi_L}\longrightarrow\,\Sigma_L\,\underset{\sigma_L}\longrightarrow\,\widetilde\Sigma_L
\end{equation}
we can associate the triangular inequality
\begin{equation}
\label{triangularE}
E_{\widehat\Sigma\Sigma}[\psi_L,\,\zeta]\,+\,E_{\Sigma\widetilde\Sigma}[\sigma_L,\,\zeta]\,\geq\,
E_{\widehat\Sigma\widetilde\Sigma}[(\sigma_L\circ \psi_L),\,\zeta]\,,
\end{equation}
where
 \begin{equation}
\label{invenfunct2tilde}
E_{\widehat\Sigma\widetilde\Sigma}[(\sigma_L\circ \psi_L),\,\zeta]\,:=\,\int_{\hat{\Sigma}_L}({\Phi}_{\widehat\Sigma\widetilde\Sigma}\,-\,1 )^2\,d\mu_{\hat{h}}\;.
\end{equation}
If $A(\widehat{\Sigma}_L)\,:=\,\int_{\widehat{\Sigma}_L}d\mu_{\hat{h}}$ and   $A({\Sigma}_L)\,:=\,\int_{{\Sigma}_L}d\mu_{{h}}$ respectively denote the area of  the surfaces $(\widehat\Sigma,\,\hat{h})$ and  $(\Sigma,\,{h})$, then we have the upper and lower bounds
\begin{equation}
\label{bounds}
A(\widehat{\Sigma}_L)\,+\,A({\Sigma}_L)\,\geq\,E_{\widehat\Sigma\Sigma}[\psi_L,\,\zeta]\,\geq\,\left( \sqrt{\A(\widehat{\Sigma}_L)}\,-\,\sqrt{A({\Sigma}_L)}  \right)^2\,.
\end{equation}
Finally,
\begin{equation}
E_{\widehat\Sigma\Sigma}[\psi_L,\,\zeta]\,=\,0
\end{equation}
iff  the surfaces $(\widehat\Sigma,\,\hat{h})$ and  $(\Sigma,\,{h})$  are isometric. 
\end{lemma}
\begin{proof}
For notational ease,  let us temporarily dismiss the action of  the linear fractional transformation
 $\zeta\,\in\,PSL(2, \mathbb{C})$ and, if there is no chance of confusion,  write  $E_{\widehat\Sigma\Sigma}[\psi_L]$ in place of the full $E_{\widehat\Sigma\Sigma}[\psi_L,\,\zeta]$. We start with proving the symmetry property 
(\ref{symmetryE}). To this end, expand the integrand in  (\ref{enfunct1}) and rewrite $E_{\widehat\Sigma\Sigma}[\psi_L]$ as
\begin{eqnarray}
\label{enfunct2}
E_{\widehat\Sigma\Sigma}[\psi_L]\,&=&\,\int_{\widehat{\Sigma}_L}( {\Phi}_{\widehat\Sigma\Sigma}\,-\,1 )^2\,d\mu_{\hat{h}}\,=\,
\int_{\widehat{\Sigma}_L}{\Phi}_{\widehat\Sigma\Sigma}^2\,d\mu_{\hat{h}}\,+\,
\int_{\widehat{\Sigma}_L}\,d\mu_{\hat{h}}\,-\,2\int_{\widehat{\Sigma}_L}{\Phi}_{\widehat\Sigma\Sigma}\,d\mu_{\hat{h}}\nonumber\\
\\
&=&\int_{\widehat{\Sigma}_L}\frac{\psi_L^*d\mu_h}{d\mu_{\hat{h}}}\,d\mu_{\hat{h}}\,+\,
A\left(\widehat{\Sigma}_L\right)\,-\,2\int_{\widehat{\Sigma}_L}{\Phi}_{\widehat\Sigma\Sigma}\,d\mu_{\hat{h}}\nonumber\\
&=&\int_{\psi_L(\widehat{\Sigma}_L)}{d\mu_h}\,+\,
A\left(\widehat{\Sigma}_L\right)\,-\,2\int_{\widehat{\Sigma}_L}{\Phi}_{\widehat\Sigma\Sigma}\,d\mu_{\hat{h}}\nonumber\\
&=&A({\Sigma}_L)\,+\,
A\left(\widehat{\Sigma}_L\right)\,-\,2\int_{\widehat{\Sigma}_L}{\Phi}_{\widehat\Sigma\Sigma}\,d\mu_{\hat{h}}\;,\nonumber
\end{eqnarray}
where we have exploited  the Radon-Nikodyn characterization of  $\widehat{\Phi}_{\widehat\Sigma\Sigma}^2$, (see (\ref{RadonNphi})), the identification $\psi(\widehat{\Sigma}_L)\,=\,\Sigma_L$,\, and the relation
\begin{equation}
\label{AreaSigma}
\int_{\widehat{\Sigma}_L}\frac{\psi_L^*d\mu_h}{d\mu_{\hat{h}}}\,d\mu_{\hat{h}}=
\int_{\widehat{\Sigma}_L}{\psi_L^*d\mu_h}=\int_{\psi(\widehat{\Sigma}_L)}{d\mu_h}=\int_{{\Sigma}_L}{d\mu_h}\,=\,A({\Sigma}_L)\;,
\end{equation}
where $A({\Sigma}_L)$ and  $A\left(\widehat{\Sigma}_L\right)$ respectively denote the area of   
$(\hat{\Sigma}_L,\,\hat{h})$ and $(\Sigma_L,\, h)$.  Along the same lines, let us compute the lightcone      comparison    functional  $E_{\Sigma\widehat\Sigma}[\psi_L^{-1}]$ associated with the inverse diffeomorphism $\psi_L^{-1}\,:\,(\Sigma_L,\, h) \,\longrightarrow\, (\hat{\Sigma}_L,\,\hat{h})$ and the corresponding conformal factor
 $\Phi_{\Sigma\widehat\Sigma}\,\in\,C^\infty({\Sigma},\,\mathbb{R})$- (see (\ref{confphiSigma})),
\begin{equation}
E_{\Sigma\widehat\Sigma}[\psi_L^{-1}]\,:=\,\int_{\Sigma_L}\left({\Phi}_{\Sigma\widehat\Sigma}\,-\,1 \right)^2\,d\mu_{h}\;.
\end{equation}
We have
\begin{equation}
\label{invE}
E_{\Sigma\widehat\Sigma}[\psi_L^{-1}]\,:=\,A\left(\widehat{\Sigma}_L\right)\,+\,A({\Sigma}_L)\,
-\,2\int_{\Sigma_L}{\Phi}_{\Sigma\widehat\Sigma}\,d\mu_{h}\;.
\end{equation}
Since     
\begin{eqnarray}
\int_{\Sigma_L}{\Phi}_{\Sigma\widehat\Sigma}\,d\mu_{h}\,&=&\,\int_{\Sigma_L}
\sqrt{\frac{d\mu_{(\psi^{-1})^*{\hat{h}}}}{d\mu_{h}}}\,
d\mu_{h}=\int_{\Sigma_L}\sqrt{\frac{d\mu_{(\psi^{-1})^*{\hat{h}}}}{d\mu_{h}}}\,\frac{d\mu_{h}}{d\mu_{(\psi^{-1})^*{\hat{h}}}}\,d\mu_{(\psi^{-1})^*{\hat{h}}}\nonumber\\
\label{conto1}\\
&=&\,\int_{\Sigma_L}\,\sqrt{\frac{d\mu_{h}}{d\mu_{(\psi^{-1})^*{\hat{h}}}}}\,(\psi^{-1})^*d\mu_{\hat{h}}\;.
\nonumber
\end{eqnarray}
On the other hand, if we take the pull back, under the action of $\psi_L^{-1}\,:\,(\Sigma_L, h)\,\longrightarrow\, (\widehat\Sigma_L, \hat{h})$, of the relation ${\Phi}_{\widehat\Sigma\Sigma}^2\,d\mu_{\hat{h}}\,=\,\psi_L^*d\mu_h$, (see (\ref{RadonNphi})), we have
\begin{equation}
\label{demu}
\left(\psi^{-1} \right)^*\left({\Phi}_{\widehat\Sigma\Sigma}^2\,d\mu_{\hat{h}}\right)\,=\,\left(\psi_L^{-1} \right)^*\left(\psi_L^*d\mu_h\right)\,
\Longrightarrow\,
{\Phi}_{\widehat\Sigma\Sigma}^2\left( \psi_L^{-1}(x) \right)\left(\left(\psi_L^{-1} \right)^*d\mu_{\hat{h}}\right)(x)\,=\,d\mu_h(x)\,,
\end{equation}
from which we get 
\begin{equation}
{\Phi}_{\widehat\Sigma\Sigma}^2\left( \psi_L^{-1}(x) \right)\,=\,\frac{d\mu_h(x)}{\left(\left(\psi_L^{-1} \right)^*d\mu_{\hat{h}}\right)(x)}\;.
\end{equation}
Hence, we can rewrite  (\ref{conto1}) as 
\begin{eqnarray}
\int_{\Sigma_L}{\Phi}_{\Sigma\widehat\Sigma}\,d\mu_{h}\,&=&\,\int_{\Sigma_L}\,
\sqrt{\frac{d\mu_{h}}{d\mu_{(\psi^{-1})^*{\hat{h}}}}}\,(\psi_L^{-1})^*d\mu_{\hat{h}}\,=\,
\int_{\Sigma_L}\,{\Phi}_{\widehat\Sigma\Sigma}\left( \psi_L^{-1} \right)\,(\psi_L^{-1})^*d\mu_{\hat{h}}\nonumber\\
&=&\,\int_{\Sigma_L}\,(\psi_L^{-1})^*\left({\Phi}_{\widehat\Sigma\Sigma}\,d\mu_{\hat{h}}\right)\,=\, 
\int_{\psi^{-1}(\Sigma_L)}\,{\Phi}_{\widehat\Sigma\Sigma}\,d\mu_{\hat{h}} \\
&=&\,
\int_{\hat{\Sigma}_L}\,{\Phi}_{\widehat\Sigma\Sigma}\,d\mu_{\hat{h}} \nonumber\;,
\end{eqnarray}
and
\begin{eqnarray}
E_{\Sigma\widehat\Sigma}[\psi^{-1}]\,&:=&\,A\left(\widehat{\Sigma}_L\right)\,+\,A({\Sigma}_L)\,
-\,2\int_{\Sigma_L}{\Phi}_{\Sigma\widehat\Sigma}\,d\mu_{h}\nonumber\\
\\
&=&\,A\left(\widehat{\Sigma}_L\right)\,+\,A({\Sigma}_L)\,
-\,2\int_{\hat{\Sigma}_L}\,{\Phi}_{\widehat\Sigma\Sigma}\,d\mu_{\hat{h}}\,=\,E_{\widehat\Sigma\Sigma}[\psi]\;.\nonumber
\end{eqnarray}
Hence, the      comparison    functional is symmetric. \\
In order to prove the triangular inequality (\ref{triangularE}) let us consider the sum
\begin{equation}
\label{sommaE}
E_{\widehat\Sigma\Sigma}[\psi_L]\,+\,E_{\Sigma\widetilde\Sigma}[\sigma_L]\,=\,
\,\int_{\hat{\Sigma}_L}({\Phi}_{\widehat\Sigma\Sigma}\,-\,1 )^2\,d\mu_{\hat{h}}\,+\,
\,\int_{{\Sigma}_L}({\Phi}_{\Sigma\widetilde\Sigma}\,-\,1 )^2\,d\mu_{{h}}\;.
\end{equation}
From the relation (\ref{demu}) we have  $d\mu_h\,=\,{\Phi}_{\widehat\Sigma\Sigma}^2\left( \psi_L^{-1} \right)\left(\psi_L^{-1} \right)^*d\mu_{\hat{h}}$, and we can write 
\begin{equation}
\int_{{\Sigma}_L}({\Phi}_{\Sigma\widetilde\Sigma}\,-\,1 )^2\,d\mu_{{h}}\,=\,
\int_{\widehat{\Sigma}_L}({\Phi}_{\Sigma\widetilde\Sigma}\,-\,1 )^2{\Phi}_{\widehat\Sigma\Sigma}^2\,d\mu_{\hat{h}}\;.
\end{equation}
Hence, 
\begin{eqnarray}
E_{\widehat\Sigma\Sigma}[\psi_L]\,+\,E_{\Sigma\widetilde\Sigma}[\sigma_L]\,&=&\,
\int_{\widehat{\Sigma}_L}\left[({\Phi}_{\widehat\Sigma\Sigma}\,-\,1 )^2\,+\,
({\Phi}_{\Sigma\widetilde\Sigma}\,-\,1 )^2{\Phi}_{\widehat\Sigma\Sigma}^2   \right]\,d\mu_{\hat{h}}\\
&\geq &\,
\int_{\widehat{\Sigma}_L}\left[({\Phi}_{\widehat\Sigma\Sigma}\,-\,1 )\,+\,
({\Phi}_{\Sigma\widetilde\Sigma}\,-\,1 ){\Phi}_{\widehat\Sigma\Sigma}   \right]^2\,d\mu_{\hat{h}}\nonumber\\
&=&\,\int_{\widehat{\Sigma}_L}({\Phi}_{\widehat\Sigma\Sigma}{\Phi}_{\Sigma\widetilde\Sigma}\,-\,1 )^2\,d\mu_{\hat{h}}\,=\,
E_{\widehat\Sigma\widetilde\Sigma}[(\sigma_L\circ \psi_L)]\nonumber\;,
\end{eqnarray}
where we have exploited the relation
\begin{equation}
{\Phi}_{\Sigma\widetilde\Sigma}\left(\psi_L \right)\,{\Phi}_{\widehat\Sigma\Sigma}\,=\,{\Phi}_{\widehat\Sigma\widetilde\Sigma}\;,
\end{equation}
which  follows from observing that the positive functions ${\Phi}_{\widehat\Sigma\widetilde\Sigma}\in\,C^\infty(\widehat\Sigma, \mathbb{R})$ and  ${\Phi}_{\Sigma\widetilde\Sigma}\in\,C^\infty(\Sigma, \mathbb{R})$  are such that
\begin{equation}
{\Phi}^2_{\widehat\Sigma\widetilde\Sigma}\,\hat{h}\,=\,\left(\sigma_L\circ \psi_L  \right)^*\tilde{h}
\,=\,\psi_L^*\left({\Phi}^2_{\Sigma\widetilde\Sigma}\,h  \right)\,=\,{\Phi}^2_{\Sigma\widetilde\Sigma}\left(\psi_L \right)\,\psi_L^*\,h\,
=\,{\Phi}^2_{\Sigma\widetilde\Sigma}\left(\psi_L \right)\,{\Phi}^2_{\widehat\Sigma\Sigma}\,\hat{h}\;,
\end{equation}
where we have set ${\Phi}_{\Sigma\widetilde\Sigma}\left(\psi_L \right)=\psi_L^*{\Phi}_{\Sigma\widetilde\Sigma}:={\Phi}_{\Sigma\widetilde\Sigma}\,\circ \,\psi_L$.\\
\\
From  (\ref{enfunct2}) and the Schwarz inequality 
\begin{equation}
\int_{\widehat{\Sigma}_L}{\Phi}_{\widehat\Sigma\Sigma}\,d\mu_{\hat{h}}\,\leq\,
\left(\int_{\widehat{\Sigma}_L}{\Phi}^2_{\widehat\Sigma\Sigma}\,d\mu_{\hat{h}}  \right)^{1/2}\,
\left(\int_{\widehat{\Sigma}_L}\,d\mu_{\hat{h}}   \right)^{1/2}\,=\,
\sqrt{A\left(\widehat{\Sigma}_L\right)A\left({\Sigma}_L\right)}\,,
\end{equation}
we get the lower bound  
\begin{eqnarray}
E_{\widehat\Sigma\Sigma}[\psi_L,\,\zeta]\,&=&\,A({\Sigma}_L)\,+\,
A\left(\widehat{\Sigma}_L\right)\,-\,2\int_{\widehat{\Sigma}_L}{\Phi}_{\widehat\Sigma\Sigma}\,d\mu_{\hat{h}}\\
&\geq &\,\,A({\Sigma}_L)\,+\,A\left(\widehat{\Sigma}_L\right)\,-\,2\sqrt{A\left(\widehat{\Sigma}_L\right)A\left({\Sigma}_L\right)}\,\nonumber\\
&=&\,   \left( \sqrt{\A(\widehat{\Sigma}_L)}\,-\,\sqrt{A({\Sigma}_L)}  \right)^2\nonumber\,,
\end{eqnarray}
where we have exploited (\ref{AreaSigma}).  The upper bound in  (\ref{bounds}) easily follows from (ref{enfunct1})
\begin{equation}
\label{upperbound}
E_{\widehat\Sigma\Sigma}[\psi_L,\,\zeta]\,:=\,\int_{\widehat{\Sigma}_L}({\Phi}_{\widehat\Sigma\Sigma}\,-\,1 )^2\,d\mu_{\hat{h}}\,
\leq\,\int_{\widehat{\Sigma}_L}{\Phi}^2_{\widehat\Sigma\Sigma}\,d\mu_{\hat{h}}\,+\,
\int_{\widehat{\Sigma}_L}\,d\mu_{\hat{h}}\,=\,A(\widehat{\Sigma}_L)\,+\,A({\Sigma}_L)\,.
\end{equation}
The proof of  the last part of the lemma follows  observing that the integrand  in $\int_{\hat{\Sigma}_L}({\Phi}_{\widehat\Sigma\Sigma}\,-\,1 )^2\,d\mu_{\hat{h}}$ is non-negative and,  as long as ${\Phi}_{\widehat\Sigma\Sigma}$ is a smooth function on $(\hat{\Sigma}_L, \hat{h})$,  the condition
\begin{equation}
\label{zeroE}
E_{\widehat\Sigma\Sigma}[\psi_L]\,=\,
\,\int_{\hat{\Sigma}_L}({\Phi}_{\widehat\Sigma\Sigma}\,-\,1 )^2\,d\mu_{\hat{h}}\,=\,0
\end{equation}
implies ${\Phi}_{\widehat\Sigma\Sigma}\,=\,1$, hence the isometry between $(\hat{\Sigma}_L, \hat{h})$  and  
$({\Sigma}_L, {h})$.  
\end{proof}
\section{A scale-dependent distance functional}
The properties of the      comparison    functional $E_{\widehat\Sigma\Sigma}[\psi_L,\,\zeta]$ indicate that we can associate with it a distance functional $d_L\left[\widehat{\Sigma}_L,\,\Sigma_L\right]$. To put the characterization of this distance in perspective, let us
recall that  a fractional linear transformations $\zeta\in\,PSL(2, \mathbb{C})$ is fully determined if, given three distinct points of $\widehat\Sigma\simeq \mathbb{S}^2$, we specify their images. We exploited this in Proposition \ref{3points}, where we  assigned  three distinct points  ${q}_{(i)}$,\,$i\,=1,2,3$ on the observer celestial sphere $({\Sigma}_L,\,{h})$, and we fixed the action of $PSL(2, \mathbb{C})$ by choosing that particular automorphism $\zeta\in\,PSL(2, \mathbb{C})$ that identifies  the inverse images $\{\psi^{-1}(q_i)\}\,\in\,\widehat{\Sigma}_L$  with three chosen points $\hat{q}_{(i)}$,\,$i\,=1,2,3$ on the reference FLRW celestial sphere  $(\widehat{\Sigma}_L,\,\hat{h})$.  Since $E_{\widehat\Sigma\Sigma}[\psi_L,\,\zeta]$ is not conformally invariant, the particular choice of the automorphism $\zeta\in\,PSL(2, \mathbb{C})$, or which is the same, the particular choice of the points  ${q}_{(i)}$,\,$i\,=1,2,3$ on $({\Sigma}_L,\,{h})$,  affects $E_{\widehat\Sigma\Sigma}[\psi_L,\,\zeta]$, hence it is natural to inquire if there is a choice of the automorphism $\zeta$ that minimizes $E_{\widehat\Sigma\Sigma}[\psi_L,\,\zeta]$.  Given the reference points $\hat{q}_{(i)}$,\,$i\,=1,2,3$ on
$(\widehat{\Sigma}_L,\,\hat{h})$, this \textit{optimal} choice for  $\zeta\in\,PSL(2, \mathbb{C})$, say $\zeta=\zeta_0$, will induce the proper selection of the alignment points ${q}_{(i)}$,\,$i\,=1,2,3$ on the observer celestial sphere $({\Sigma}_L,\,{h})$ 
by setting ${q}_{(i)}\,:=\,\psi_L\left(\zeta_0(\hat{q}_{(i)})\right)$. In order to characterize this optimal choice,  we need to minimize 
$E_{\widehat\Sigma\Sigma}[\psi_L,\,\zeta]$ over a suitable class of functions, and a natural strategy, according to these remarks, is to keep fixed the diffeomorphism\footnote{The reason for keeping $\psi_L$ fixed is directly related to the fact that $\psi_L$ it is constructed by using the null geodesics along the past lightcones by exploiting the exponential maps  (see (\ref{psi0}) ), and this is the way actual cosmological observations are carried out.} $\psi_L$ as well the points $\hat{q}_{(i)}$,\,$i\,=1,2,3$ on the reference  $(\widehat{\Sigma}_L,\,\hat{h})$, and let vary in a controlled way the automorphism $\zeta\in\,PSL(2, \mathbb{C})$, so as to minimize 
$E_{\widehat\Sigma\Sigma}[\psi_L,\,\zeta]$.  We also need a slightly more general setting that will allow us to deal  with  celestial spheres
$(\Sigma_L,\, h) $ on a lightcone region $\mathcal{C}^-_{L}(p,g)$ where caustics develop (hence, relaxing in a controlled way  the regularity of $\psi_L$ allowing for exponential mappings which are no longer injective).  In other words, we need to  extend  $\psi_L\circ\zeta\,:\,\widehat{\Sigma}_L\longrightarrow\Sigma_L$ to be a member of  a more  general space of maps which allow for 
the low regularity setting associated with the possible presence of (isolated) caustics. We start with a more precise characterization of the Sobolev space of maps $W^{1,2}(\widehat\Sigma,\,\Sigma)$, mentioned on passing in commenting Definition \ref{sepfunct}.  To define $W^{1,2}(\widehat\Sigma,\,\Sigma)$ we follow a standard approach in harmonic map theory and use Nash embedding theorem \cite{GuntNash}, \cite{Schwartz}, by considering the compact surface  $(\Sigma_L, h)$  isometrically embedded into some Euclidean space $\mathbb{E}^m\,:=\,(\mathbb{R}^m, \delta )$ for $m$ sufficiently large. In particular,  if $J: (\Sigma_L, h)\hookrightarrow \mathbb{E}^m$ is any such an embedding  then  we define the Sobolev space of maps
\begin{equation}
{W}^{1, 2}_{(J)}(\widehat\Sigma, \Sigma)\, :=\,\{\phi\in {W}^{1, 2}(\widehat\Sigma,\,\mathbb{R}^m)\left.\right|\,\phi(\hat\Sigma_L)\subset J(\Sigma_L) \}\;,
\end{equation} 
where ${W}^{1, 2}(\widehat\Sigma,\mathbb{R}^m)$ is the Hilbert space of square summable  $\varphi:\widehat\Sigma \rightarrow \mathbb{R}^m$, with (first) distributional derivatives in  $L^2(\widehat\Sigma,\mathbb{R}^m)$, endowed with the norm
\begin{equation}
 \parallel \phi \parallel_{{W}^{1,2}}\,:=\,\int_{\widehat\Sigma}\,\left(\phi^a(x)\,\phi^b(x)\,\delta _{ab}\,+\,
\hat{h}^{\mu\nu}(x)\,\frac{\partial \phi^{a}(x)}{\partial x^{\mu}}
\frac{\partial\phi^{b}(x)}{\partial x^{\nu}}\,\delta _{ab}  \right)\,d\mu_{\hat{h}}\;,
\end{equation}
where, for $\phi(x)\in\,J(\Sigma_L)\subset  \mathbb{R}^m$,\, $a,b=1,\ldots,m$ label coordinates in $(\mathbb{R}^m,\,\delta)$, and $d\mu_{\hat{h}}$ denotes the Riemannian measure on $(\widehat\Sigma, \hat{h})$. This characterization is independent of $J$ since  $\Sigma_L$ is compact, and  in that case for any two isometric embeddings $J_1$ and $J_2$, the corresponding spaces of maps  ${W}^{1, 2}_{(J_1)}(\widehat\Sigma, \Sigma)$ and ${W}^{1, 2}_{(J_2)}(\widehat\Sigma, \Sigma)$ are homeomorphic \cite{helein}. For this reason, in what follows we shall simply write ${W}^{1, 2}(\widehat\Sigma, \Sigma)$. 
The set of maps ${W}^{1, 2}(\widehat\Sigma, \Sigma)$  provides  the minimal regularity allowing for the characterization of the energy functional  $E_{\widehat\Sigma\Sigma}[\psi_L,\,\zeta]$.  Maps of class ${W}^{1, 2}(\widehat\Sigma, \Sigma)$ are not necessarily  continuous and, even if   
the space of smooth maps  ${C}^{\infty}(\widehat\Sigma, \Sigma)$ is dense \cite{schoen} in 
${W}^{1, 2}(\widehat\Sigma, \Sigma)$, to carry out explicit computations, in what follows we must further require that $\phi \in {W}^{1, 2}(\widehat\Sigma, \Sigma)$ is localizable  (cf.  \cite{jost}, Sect. 8.4) and keeps track both of the given $\psi_L$ and of  the   three \textit{alignment} points $\hat{q}_{(i)}$,\,$i\,=1,2,3$ on  $(\widehat{\Sigma}_L,\,\hat{h})$. The only freedom remaining is in the conformal group automorphisms  $\zeta\in\,PSL(2, \mathbb{C})$ acting on  $(\widehat{\Sigma}_L,\,\hat{h})$, and in terms of which
we need to control that the  images of   the reference points $\hat{q}_{(i)}\in\widehat{\Sigma}_L$ stay separated and do not concentrate in a small neighborhood of  ${\Sigma}_L$. Hence, and  for  a fixed $\psi_L\,\in\,{W}^{1, 2}(\widehat\Sigma, \Sigma)$,  we define the  space of maps over which  $E_{\widehat\Sigma\Sigma}[\psi_L,\,\zeta]$   is minimized according to 
 the following definition.
\begin{defi}
\label{FunctSpace}
Let us assume that $\psi_L\,:\,(\widehat{\Sigma}_L,\,\hat{h})\,\longrightarrow\,(\Sigma_L,\, h) $ is, for almost all points of  $\widehat{\Sigma}_L$,  a  ${W}^{1, 2}(\widehat\Sigma, \Sigma)$ diffeomorphism between the  two celestial spheres\footnote{Hence, we are assuming that there can be a finite collections of points for which the exponential mapping $\exp_p$ along $\mathcal{C}^-_{L}(p,g)$ may be not injective.},  and let  
$\hat{q}_{(i)}$,\,$i\,=1,2,3$,\, be the three distinguished points on  
$(\widehat{\Sigma}_L,\,\hat{h})$, characterizing the three reference past-directed null directions on $\mathcal{C}^-_{L}(p,\hat{g})$ introduced in Proposition \ref{3points}.   A map $\phi\,:=\,\psi_L\circ\zeta\,\in\,{W}^{1, 2}(\widehat\Sigma, \Sigma)$, with $\zeta\in\,PSL(2, \mathbb{C})$,\, is said to be $\epsilon$-localizable if: (i) For every $\hat{q} \in \widehat\Sigma_L$ there exists a metric disks
$D(\hat{q},\,\delta  ):=\{y\in \widehat\Sigma_L\,|\,d_{\gamma}(\hat{q},y)\,\leq \,\delta\} \subset \widehat\Sigma_L$, of radius $\delta>0$, with smooth boundary $\partial\,D$, and containing at most 
one of the three points $\hat{q}_{(i)}$,\,$i\,=1,2,3$; and (ii) Corresponding to each of these disks, there exists
 a metric disk  $B(q,\,\epsilon)\,=\,\phi(D(\hat{q},\,\delta  ))\,:=\,\{x\in \Sigma_L\,\,|\,d_h(q, x)\leq  \epsilon\}\,\subset\, (\Sigma_L, h)$  centered at $\phi(x_0)\,:=\,q\,\in\, M$, of radius $r>0$ such that $\phi(D(x_0,\,\delta  ))\subset B(q,\,r)$, with  $\phi(\partial D)\,\subset\,B(q,\,\epsilon)$.
 Under such assumptions,  we consider, for fixed $\psi_L$,  the space of maps
 \begin{eqnarray}
 &&\mathrm{Map}_\psi(\widehat\Sigma_L, \Sigma_L)\,:=\,\left\{\phi\,:=\,\psi_L\circ\zeta\,\in\,{W}^{1, 2}(\widehat\Sigma, \Sigma)\cap\,{C}^{0}(\widehat\Sigma, \Sigma)\,,\,\,\right.\\
&&\left. \left.\zeta\in\,PSL(2, \mathbb{C})\,\,\,\right|\,\,\phi\,:=\,\psi_L\circ\zeta\,\,\,\,\mathrm{is\,\, \epsilon-localizable\,\,and}\,\,\Phi_{\widehat\Sigma\Sigma}(\hat{q})\geq0   \right\}\;,\nonumber
 \end{eqnarray}
where the non-negativity requirement $\Phi_{\widehat\Sigma\Sigma}(\hat{q})\geq0$ is assumed to hold for almost all points of $\widehat\Sigma_L$.
 \end{defi}
 As in harmonic map theory, there is a further delicate issue related to the fact that maps in $\mathrm{Map}_\psi(\widehat\Sigma_L, \Sigma_L)$ are partitioned in different homotopy classes.  Recall that every map from $\mathbb{S}^2$ into itself  is characterized by the degree of the map \cite{Petersen}, measuring how many times the map wraps  $\mathbb{S}^2$ around itself. In particular through the action of  a sequence of conformal dilations  $\in\,PSL(2, \mathbb{C})$ of the form $\zeta\longmapsto \zeta'_{(k)}\,:=\,\omega_{(k)}\zeta$ where $\omega_{(k)}\in\mathbb{R}$  we can easily construct sequences of mappings $\{\varphi_{(k)}\}$ that tend to focus all points of  a disk $D$ in $\mathbb{S}^2$ toward a given point (say the north pole). Physically this corresponds to the effect of acting with a sequence of Lorentz boosts (with rapidity $\log\,\omega_{(k)}$) on an observer $P$ who is looking at the giving region $D$ of the celestial sphere. From the point of view of $P$, one can also interpret this as a focusing of the past null geodesics eventually leading to the formation of a caustic point. Regardless of the physical interpretation, in harmonic map theory this sort of behavior leads to the phenomenon of  \textit{bubble convergence} when discussing the minimization problem for  the harmonic map energy functional \cite{Uhlenbeck}. In our case, we can exploit the analogous of bubbling convergence to our advantage in order to extend our analysis to the case when caustics are present.  We can model the generation of caustic points as the results of  a  focusing mapping,  such as the $\{\varphi_{(k)}\}$ described above,  converging in $\mathrm{Map}_\psi(\widehat\Sigma_L, \Sigma_L)$,     to a $\mathrm{deg}\,\varphi\,=\,h\,>\,0$ map (if there are $h-1$ caustic points in $(\Sigma_L, h)$). We postpone the details of such analysis to a paper in preparation \cite{CarFam},  and limit here our analysis to show that  $E_{\widehat\Sigma\Sigma}[\psi_L,\,\zeta]$ can be minimized over diffeomorphisms in  $\mathrm{Map}_\psi(\widehat\Sigma_L, \Sigma_L)$.
\begin{theorem} (The lightcone comparison distance at scale $L$).
\label{distheorem}
The functional $E_{\widehat\Sigma\Sigma}[\psi_L,\,\zeta]$  achieves a minimum on  $\mathrm{Map}_\psi(\widehat\Sigma_L, \Sigma_L)$, and 
\begin{equation}
d_L\left[\widehat{\Sigma}_L,\,\Sigma_L\right]\,:=\,\inf_{\psi_L\circ\zeta\in \mathrm{Map}_\psi(\widehat\Sigma_L, \Sigma_L)}\,E_{\widehat\Sigma\Sigma}[\psi_L,\,\zeta]
\end{equation}
defines a scale-dependent   distance between the celestial  spheres  $(\widehat{\Sigma}_L,\,\hat{h})$ and $(\Sigma_L,\, h) $ on the lightcone regions 
$\mathcal{C}^-_{L}(p,\hat{g})$ and $\mathcal{C}^-_{L}(p,g)$.
\end{theorem}
\begin{proof}
To simplify notation let us set $\phi\,:=\,\psi_L\circ\zeta$. Since we have the upper bound $E_{\widehat\Sigma\Sigma}[\phi] \leq\,C_L\,:=\,A(\widehat{\Sigma}_L)\,+\,A({\Sigma}_L)$, (see (\ref{bounds})),  we can limit our analysis to the subset of  maps 
\begin{equation}
\mathrm{Map}_{\psi, C_L}(\widehat\Sigma_L, \Sigma_L) \,:=\,\left\{\phi\,\in\,\mathrm{Map}_{\psi}(\widehat\Sigma_L, \Sigma_L)\,\,|\,\,\mathrm{s.t.}\,\,
 E_{\widehat\Sigma\Sigma}[\phi] \,\leq\,C_L \right\}\;.
\end{equation}
According to  Definition  \ref{FunctSpace}, the space of maps  $\mathrm{Map}_{\epsilon, C_L}(\widehat\Sigma, \Sigma) $ is equicontinuous, namely, for any point $\hat{q}\in\,\hat\Sigma$ we can choose the disk $D(\hat{q}_1, \delta)$ (for notation see Definition  \ref{FunctSpace}) in such a way that for a given $\epsilon\,>\,0$, \,\,$\phi(\hat{q}_1)$ and 
$\phi(\hat{q}_2)$ are such that $d_h\left(\phi(\hat{q}_1),\,\phi(\hat{q}_2) \right)\,<\,\epsilon$, for all $\hat{q}_2\,\in\,D(\hat{q}_1, \delta)$ and all $\phi\,\in\,\mathrm{Map}_{\psi, C_L}(\widehat\Sigma_L, \Sigma_L)$. Hence, a minimizing sequence $\{\phi_{(k)} \}_{k\in\mathbb{N}}\,\in\,\mathrm{Map}_{\epsilon, C_L}(\widehat\Sigma, \Sigma)$ for  $E_{\widehat\Sigma\Sigma}[\phi] $ is equicontinuous. By selecting a subsequence we may assume that 
$\{\phi_{(k)} \}$ converges to a continuous map $\varphi$ which is also the weak limit of  $\{\phi_{(k)} \}$ in $W^{1, 2}(\widehat\Sigma, \Sigma)$, since this latter is a weakly compact space of maps.\\
Since
\begin{equation}
\label{EAA}
E_{\widehat\Sigma\Sigma}[\phi]\,=\,A({\Sigma}_L)\,+\,
A\left(\widehat{\Sigma}_L\right)\,-\,2\int_{\widehat{\Sigma}_L}{\Phi}_{\widehat\Sigma\Sigma}(\phi)\,d\mu_{\hat{h}}\,,
\end{equation}
a minimizing (sub)sequence $\{\phi_{(k)} \}$  for  $E_{\widehat\Sigma\Sigma}[\phi]$ corresponds to a maximize sequence for the functional
$\int_{\widehat{\Sigma}_L}{\Phi}_{\widehat\Sigma\Sigma}(\phi)\,d\mu_{\hat{h}}$.  Hence,  given $\delta\,>\,0$,\, there exists $k_0$ such that for all $k\,\geq\,k_0$, we have
\begin{equation}
\label{deltaE}
\int_{\widehat{\Sigma}_L}{\Phi}_{\widehat\Sigma\Sigma}(\overline\phi)\,d\mu_{\hat{h}}\,\geq\,
\int_{\widehat{\Sigma}_L}{\Phi}_{\widehat\Sigma\Sigma}(\phi_{(k)})\,d\mu_{\hat{h}}\,-\delta\\,,
\end{equation}
along a minimizing sequence $\{\phi_{(k)} \}\,\longrightarrow\,\overline\phi$  for  the functional $E_{\widehat\Sigma\Sigma}[\phi]$, and 
where ${\Phi}_{\widehat\Sigma\Sigma}(\phi_{(k)})$ is non-negative  for almost all points of $\widehat\Sigma_L$. By adding and subtracting 
$\int_{\widehat{\Sigma}_L}{\Phi}_{\widehat\Sigma\Sigma}(\phi_{(k)})\,d\mu_{\hat{h}}$ to (\ref{EAA}), (evaluated for $\overline\phi$), and by taking into account  (\ref{deltaE}),  we get
\begin{equation}
\label{plsemi}
E_{\widehat\Sigma\Sigma}[\overline\phi]\,=\, E_{\widehat\Sigma\Sigma}[\phi_{(k)}]\,-\,
2\int_{\widehat{\Sigma}_L}\left({\Phi}_{\widehat\Sigma\Sigma}(\overline\phi)\,-\, {\Phi}_{\widehat\Sigma\Sigma}(\phi_{(k)})  \right)\,d\mu_{\hat{h}}\,\leq\,E_{\widehat\Sigma\Sigma}[\phi_{(k)}]\,+\,2\delta\;,
\end{equation}
for all $k\,\geq\,k_0$.  Since the choice of $\delta\,>\,0$ is arbitrary, (\ref{plsemi}) implies  that the functional $E_{\widehat\Sigma\Sigma}[\phi]$ is lower semicontinuous,  \textit{i.e.},
\begin{equation}
E_{\widehat\Sigma\Sigma}[\overline\phi]\,\leq\,\lim_k\inf\,E_{\widehat\Sigma\Sigma}[\phi_k]  
\end{equation}
for all $\varphi \in\,\mathrm{Map}_{\epsilon, C_L}(\widehat\Sigma, \Sigma)$ with $\phi_k$ weakly converging, in the  above sense,  to $\overline\phi$.   Hence,  $\{\phi_{(k)} \}\,\longrightarrow\,\overline\phi$ minimizes  $E_{\widehat\Sigma\Sigma}[\phi]$ in the space of maps $\mathrm{Map}_{\epsilon}(\widehat\Sigma, \Sigma)$, as stated.\\
If we set
\begin{equation}
d_L\left[\widehat{\Sigma}_L,\,\Sigma_L\right]\,:=\,\inf_{\psi_L\circ\zeta\in \mathrm{Map}_\epsilon(\widehat\Sigma, \Sigma)}\,E_{\widehat\Sigma\Sigma}[\psi_L,\,\zeta]
\end{equation}
then as a consequence of the properties of the functional 
$E_{\widehat\Sigma\Sigma}[\phi]$, described in  Lemma \ref{propfunct}, we have that $d_L\left[\widehat{\Sigma}_L,\,\Sigma_L\right]$ provides a scale dependent distance function between the physical celestial sphere $({\Sigma}_L,\,h)$ and the reference FLRW celestial 
sphere $(\widehat{\Sigma}_L,\,\hat{h})$, as the scale $L$ varies. In particular, with the notation of  Lemma \ref{propfunct} we have \textit{(i)\,Non-negativity}\,\, 
$d_L\left[\widehat{\Sigma}_L,\,\Sigma_L  \right]\,\geq\,0$;\, \textit{(ii)}\,\,  $d_L\left[\widehat{\Sigma}_L,\,\Sigma_L \right]\,=\,0$\,\,iff  $(\widehat{\Sigma}_L,\,\hat{h})$ and $(\Sigma_L,\, h)$ are isometric;\,\textit{(iii)\, Symmetry }\,\, $d_L\left[\widehat{\Sigma}_L,\,\Sigma_L  \right]=
d_L\left[{\Sigma}_L,\,\widehat{\Sigma}_L  \right]$;\,\textit{(iv)\, Triangular inequality}\,\, 
$d_L\left[\widehat{\Sigma}_L,\,\widetilde{\Sigma}_L \right]\leq
d_L\left[\hat{\Sigma}_L,\,\Sigma_L  \right]+d_L\left[{\Sigma}_L,\,\widetilde{\Sigma}_L \right]$.
\end{proof}
\section{The physical meaning of  $d_L\left[{\widehat\Sigma}_L,\,{\Sigma}_L  \right]$}
The distance functional  $d_L\left[\widehat{\Sigma}_L,\,{\Sigma}_L  \right]$  is a geometric quantity that we can associate with the observer  who wish to describe with a Friedmannian bias the cosmological region where inhomogeneities may dominate. To appreciate what this role implies, let us briefly discuss the physical interpretation of $d_L\left[\widehat{\Sigma}_L,\,{\Sigma}_L  \right]$, when we probe the light cone regions  $\mathcal{C}^-_{L}(p,\hat{g})$ and $\mathcal{C}^-_{L}(p,g)$ over a sufficiently small length scale $L$.
 If $\bar{\phi}$ denotes the minimizing map  characterized in Theorem \ref{distheorem}, we can write
\begin{equation}
\label{intdist}
d_L\left[\widehat{\Sigma}_L,\,{\Sigma}_L  \right]\,=\,E_{\widehat\Sigma\Sigma}[\bar\phi]\,:=\,A\left(\widehat{\Sigma}_L\right)\,+\,A({\Sigma}_L)\,
-\,2\int_{\widehat\Sigma_L}{\Phi}_{\widehat\Sigma\Sigma}\,d\mu_{\hat{h}}\,.
\end{equation} 
To simplify matters, we assume that at the given length scale $L$ the corresponding region $\mathcal{C}^-_{L}(p,g)$ is caustic free, and  parametrize  ${\Phi}_{\Sigma\widehat\Sigma}(\bar\phi)$ as
\begin{equation}
{\Phi}_{\Sigma\widehat\Sigma}(\bar\phi)\,=\,1\,+\,F(\bar\phi)\;,
\end{equation}
where $F(\bar\phi)$ is a smooth function (not necessarily positive) which, by discarding the traceless lensing shear, may be thought of as describing  the (small) local isotropic focusing distortion of the images of the astrophysical sources on $(\Sigma, h)$  due to gravitational lensing, (see the expression (\ref{etametric0}) of the sky-mapping metric $h$).
 Under these assumptions, we can write  (\ref{intdist}) as 
\begin{equation}
\label{distapprox0}
d_L\left[\widehat{\Sigma}_L,\,{\Sigma}_L  \right]\,=\,A\left({\Sigma}_L\right)\,-\,A(\widehat{\Sigma}_L)\,
-\,2\int_{\widehat\Sigma_L}\,F(\bar\phi)\,d\mu_{\hat{h}}\,.
\end{equation} 
This expression can be further specialized if  we exploit the asymptotic expressions of the area $A\left(\widehat{\Sigma}_L\right)$ and 
$A\left({\Sigma}_L\right)$ of the two surfaces $(\widehat{\Sigma}_L, \,\hat{h})$, \,$({\Sigma}_L, \,{h})$ on the corresponding lightcones
$\mathcal{C}^-_{{L}}(p, \hat{g})$ and  $\mathcal{C}^-_{{L}}(p, {g})$. These asymptotic expressions can be obtained if we consider the associated causal past regions $\mathcal{J}^-_{{L}}(p, \hat{g})$  and $\mathcal{J}^-_{{L}}(p, {g})$ sufficiently near the (common) observation point $p$, in particular when the length scale $L$ we are probing is small with respect to the "cosmological" curvature scale.  Under such assumption, there is a unique maximal 3-dimensional region $V_L^3(p)$, embedded in   $\mathcal{J}^-_{{L}}(p, {g})$, having the surface  $({\Sigma}_L, \,{h})$ as its boundary. This surface intersects the world line $\gamma(\tau)$ of the observer $p$ at the point $q=\gamma(\tau_0\,=\,-\,L)$ defined by the given length scale $L$. For the reference FLRW the analogous set up is associated to the constant-time slicing of the FLRW spacetime $(M, \hat{g})$ considered. The corresponding  3-dimensional region $\widehat{V}_L^3(p)$, embedded in   $\mathcal{J}^-_{{L}}(p, \hat{g})$, has the surface  $(\widehat{\Sigma}_L, \,\hat{h})$ as its boundary.  The FLRW observer $\hat{\gamma}(\hat\tau)$ will intersect $\widehat{V}_L^3(p)$ at the point $\hat{q}=\hat{\gamma}(\hat{\tau}_0\,=\,-\,L)$.  By introducing  geodesic normal coordinates $\{X^i\}$ in $\mathcal{J}^-_{{L}}(p, {g})$  and  $\{Y^k\}$ in $\mathcal{J}^-_{{L}}(p, \hat{g})$, respectively based at the point $q$ and $\hat{q}$, we can pull back the  metric tensors $g$ and $\hat{g}$ to $T_{q}M$ and $T_{\hat{q}}M$, and obtain the classical normal coordinate development of the metrics
$g$ and $\hat{g}$ valid in a sufficiently small convex neighborhood of  $q$ and $\hat{q}$. Explicitly, for the (more relevant case of the)  metric $g$, we have   (see \textit{e. g.} Lemma 3.4 (p. 210) of  \cite{SchoenYau} or \cite{Petersen}) 
\begin{eqnarray}
&&\left((exp_q)^*\,g \right)_{ef}\,=\,\eta_{ef}\,-\,\frac{1}{3}\,\mathrm{R}_{eabf}|_qX^aX^b\,-\,\frac{1}{6}\,\nabla_c\mathrm{R}_{eabf}|_qX^aX^bX^c\nonumber\\
\nonumber\\
&&+\,\left(-\,\frac{1}{20}\,\nabla_c\nabla_d\mathrm{R}_{eabf}\,+\,\frac{2}{45}\,\mathrm{R}_{eabm}\,\mathrm{R}^m_{fcd} \right)_q\,X^aX^bX^cX^d\,+\,\ldots\nonumber\;,
\end{eqnarray}
where $\mathrm{R}_{abcd}$ is the Riemann tensor of the metric $g$ (evaluated at the point 
$q$).  The induced expansion in the pulled-back Lorentzian measure $\left((exp_{s(\eta)})^*d\mu_{g}\right)$ and a rather delicate analysis (related to the spacetime geometry of  \textit{causal diamonds})  described (at various levels of rigor) in \cite{Berthiere}, \cite{GibbSol1}, \cite{GibbSol2}, \cite{Myrheim})   provides, to leading order in $L$, the following expressions for the area  of  $({\Sigma}_L, \,{h})$ and  
$(\widehat{\Sigma}_L, \,\hat{h})$, 
\begin{equation}
\label{Areaasymp1}
A\left({\Sigma}_L\right)\,=\,{\pi}\,L^2\,\left(1\,-\,\frac{1}{72}\,L^2\,\mathrm{R}(q)\,+\,\ldots  \right)\;,
\end{equation}
and
\begin{equation}
\label{Areaasymp1}
A\left(\widehat{\Sigma}_L\right)\,=\,{\pi}\,L^2\,\left(1\,-\,\frac{1}{72}\,L^2\,\widehat{\mathrm{R}}(\hat{q})\,+\,\ldots  \right)\;,
\end{equation}
Introducing these expressions in (\ref{distapprox0}) we get
\begin{equation}
\label{distapprox}
d_L\left[\widehat{\Sigma}_L,\,{\Sigma}_L  \right]\,=\,\frac{\pi}{72}\,L^4\,\left(\widehat{\mathrm{R}}(\hat{q})\,-\,{\mathrm{R}}({q}) \right)\,
-\,2\int_{\widehat\Sigma_L}\,F(\bar\phi)\,d\mu_{\hat{h}}\,+\,\ldots\,.
\end{equation} 
We can rewrite this equivalently as
\begin{equation}
\label{Rdapprox}
\widehat{\mathrm{R}}(\hat{q})\,=\,{\mathrm{R}}({q})\,+\,\frac{72}{\pi}\frac{d_L\left[\widehat{\Sigma}_L,\,{\Sigma}_L  \right]}{L^4}\,
+\,\frac{144}{\pi L^4}\,\int_{\widehat\Sigma_L}\,F(\bar\phi)\,d\mu_{\hat{h}}\,+\,\ldots\,.
\end{equation}
The asymptotics   (\ref{distapprox}), (but also the very characterization of the distance functional  $d_L\left[\widehat{\Sigma}_L,\,{\Sigma}_L  \right]$), shows clearly that the lightcone comparison functional $E_{\widehat\Sigma \Sigma}$ and the associated distance $d_L\left[\widehat{\Sigma}_L,\,{\Sigma}_L  \right]$ provide a  generalization of the lightcone theorem 
\cite{ChoquetLightCone} proved by Y. Choquet-Bruhat, P. T. Chrusciel, and  J. M. Martin-Garcia. The normal coordinates asymptotics  
(\ref{Rdapprox}) is also interesting since it directly connects  $d_L\left[\widehat{\Sigma}_L,\,{\Sigma}_L  \right]$ to the fluctuations in the spacetime scalar curvature: if we decide to keep on in modeling with a FLRW solution a cosmological spacetime, homogeneous on large scale but highly inhomogeneous at smaller scale, then the associated scalar curvature    $\widehat{\mathrm{R}}(\hat{q})$ can be approximately identified with the physical scalar curvature ${\mathrm{R}}(\hat{q})$, with a rigorous level of scale dependence precision, only if we take into account the contribution provided by  the lightcone distance functional $d_L\left[\widehat{\Sigma}_L,\,\Sigma_L  \right]$ (and by the average of the local focusing  term $F(\bar\phi)$). This can be of some interest in addressing  backreaction problems in cosmology (see \textit{e.g.}, \cite{Buchert}).  Finally,  $d_L\left[\widehat{\Sigma}_L,\,{\Sigma}_L  \right]$ can be also of some use in providing a rigorous way of addressing some aspect of the best-fitting problem in cosmology (see \cite{EllisPhRep} and \cite{Ellis2}), roughly speaking, the strategy is to vary the family of model spacetimes $(M, \hat{g})$ (for instance, the family of FLRW solutions, or the larger family of  homogeneous spacetimes) in such a way to minimize (over the relevant interval of length scales $L$) the distance functional  $d_L\left[\widehat{\Sigma}_L,\,{\Sigma}_L  \right]$ between the physical celestial spheres $(\widehat{\Sigma}_L, \hat{h})$ and the family of reference celestial spheres $(\widehat{\Sigma}_L, \hat{h})$ associated with the  model spacetimes $(M, \hat{g})$ adopted. 

\section{Acknowledgments}
We wish to thank Thomas Buchert and Dennis Stock for valuable discussions.   This work has been partially supported by
the European project ERC-2016-ADG advanced grant "arthUS": \textit{advances in the research on theories of the dark Universe}.

\end{document}